\documentclass[aps,prl,reprint,amsmath,amssymb,superscriptaddress,showpacs]{revtex4-2}
\pdfoutput=1
\usepackage{graphicx}
\usepackage{dcolumn}
\usepackage{bm}
\usepackage[colorlinks,linkcolor=blue,anchorcolor=blue,citecolor=blue,urlcolor=blue,filecolor=blue,menucolor=blue,runcolor=blue]{hyperref}

\begin{document}
\title{ARPES signature of the competition between magnetic order and Kondo effect in \texorpdfstring{${\mathbf{CeCoGe}}_{3}$}{}}
\author{Peng Li}
\affiliation{Center for Correlated Matter and Department of Physics, Zhejiang University, Hangzhou 310058, China}
\author{Huiqing Ye}
\affiliation{Center for Correlated Matter and Department of Physics, Zhejiang University, Hangzhou 310058, China}
\author{Yong Hu}
\affiliation{Paul Scherrer Institute, Swiss Light Source, CH-5232 Villigen PSI, Switzerland}
\author{Yuan Fang}
\affiliation{Center for Correlated Matter and Department of Physics, Zhejiang University, Hangzhou 310058, China}
\author{Zhiguang Xiao}
\affiliation{Center for Correlated Matter and Department of Physics, Zhejiang University, Hangzhou 310058, China}
\author{Zhongzheng Wu}
\affiliation{Center for Correlated Matter and Department of Physics, Zhejiang University, Hangzhou 310058, China}
\author{Zhaoyang Shan}
\affiliation{Center for Correlated Matter and Department of Physics, Zhejiang University, Hangzhou 310058, China}
\author{Ravi P. Singh}
\affiliation{Department of Physics, Indian Institute of Science Education and Research Bhopal, Bhopal 462066, India}
\author{Geetha Balakrishnan}
\affiliation{Department of Physics, University of Warwick, Coventry CV4 7AL, United Kingdom}
\author{Dawei Shen}
\affiliation{State Key Laboratory of Functional Materials for Informatics and Center for Excellence in Superconducting Electronics, SIMIT, Chinese Academy of Science, Shanghai, China}
\author{Yi-feng Yang}
\affiliation{Beijing National Laboratory for Condensed Matter Physics, Institute of Physics, Chinese Academy of Science, Beijing 100190, China}
\author{Chao Cao}
\affiliation{Center for Correlated Matter and Department of Physics, Zhejiang University, Hangzhou 310058, China}
\author{Nicholas C. Plumb}
\affiliation{Paul Scherrer Institute, Swiss Light Source, CH-5232 Villigen PSI, Switzerland}
\author{Michael Smidman}
\email {msmidman@zju.edu.cn}
\affiliation{Center for Correlated Matter and Department of Physics, Zhejiang University, Hangzhou 310058, China}
\author{Ming Shi}
\affiliation{Paul Scherrer Institute, Swiss Light Source, CH-5232 Villigen PSI, Switzerland}
\author{Johann Kroha}
\affiliation{Physikalisches Institut, Universit\"at Bonn, Nussallee 12, 53115 Bonn, Germany}
\author{Huiqiu Yuan}
\email {hqyuan@zju.edu.cn}
\affiliation{Center for Correlated Matter and Department of Physics, Zhejiang University, Hangzhou 310058, China}
\affiliation{Collaborative Innovation Center of Advanced Microstructures, Nanjing University, Nanjing, China}
\affiliation{State Key Laboratory of Silicon Materials, Zhejiang University, Hangzhou 310058, China}
\author{Frank Steglich}
\affiliation{Center for Correlated Matter and Department of Physics, Zhejiang University, Hangzhou 310058, China}
\affiliation{Max Planck Institute for Chemical Physics of Solids, 01187 Dresden, Germany}
\author{Yang Liu}
\email {yangliuphys@zju.edu.cn}
\affiliation{Center for Correlated Matter and Department of Physics, Zhejiang University, Hangzhou 310058, China}
\affiliation{Collaborative Innovation Center of Advanced Microstructures, Nanjing University, Nanjing, China}
\date{\today}%
\addcontentsline{toc}{chapter}{Abstract}

\begin{abstract}
The competition between magnetic order and Kondo effect is essential for the rich physics  of heavy fermion systems. Nevertheless, how such competition is manifested in the quasiparticle bands in a real periodic lattice remains elusive in spectroscopic experiments. Here we report a high-resolution photoemission study of the antiferromagnetic Kondo lattice system CeCoGe$_3$ with a high $T_{N1}$ of 21 K. Our measurements reveal a weakly dispersive $4f$ band at the Fermi level near the \textit{Z} point, arising from moderate Kondo effect. The intensity of this heavy $4f$ band exhibits a logarithmic increase with lowering temperature and begins to deviate from this Kondo-like behavior below $\sim$25 K, just above $T_{\rm N1}$, and eventually ceases to grow below $\sim$12 K. Our work provides direct spectroscopic evidence for the competition between magnetic order and the Kondo effect in a Kondo lattice system with local-moment antiferromagnetism, indicating a distinct scenario for the microscopic coexistence and competition of these phenomena, which might be related to the real-space modulation.
\end{abstract}

\maketitle

Heavy-fermion (HF) compounds are prototypical strongly correlated electron systems and host many intriguing electronic phases, including Fermi liquids with extremely heavy quasiparticles, quantum criticality, non-Fermi-liquid behavior, and unconventional superconductivity \cite{Coleman2007heavy,gegenwart2008quantum,Lohneysen2007Fermi,Sachdev2011Quantum,Steglich1979Superconductivity,Mathur1998Magnetically,Stewart2001NFL}. Lying at the heart of the rich physics in HF systems is the competition between the Ruderman-Kittel-Kasuya-Yosida (RKKY) magnetic exchange interaction \cite{RK,K,Y} and the many-body Kondo effect \cite{Coleman2007heavy,Kondo1964,Hewson1993Kondo,Nejati2017,kirchner2020colloquium,Yang2012PNAS}. Although both interactions stem from the same spin-exchange coupling $J$ between the conduction electrons and local moments, the strengths of these interactions exhibit distinct dependences on $J$, leading to a magnetically ordered (paramagnetic) ground state at small (large) $J$. Their competition is manifested in the Doniach phase diagram \cite{Doniach1977Kondo}, which underlies the phase diagrams of many HF compounds upon tuning with parameters such as pressure or chemical doping.

Of particular interest is the nature of the $4f$ electrons that underlies the aforementioned competition in HF systems with different antiferromagnetic (AFM) characters, as well as its connection with quantum criticality. In itinerant AFM systems such as CeCu$_2$Si$_2$ \cite{Zwicknagl1993Physica,Gegenwart1998,Stockert2004,Arndt2011,Wu2021CeCu2Si2}, the magnetic order is mainly driven by nesting of the renormalized Fermi surface (FS) and the associated quantum critical point (QCP) is of the HF spin-density-wave (SDW) type \cite{Gegenwart1998,Arndt2011}. On the other hand, local-moment AFM order is primarily mediated by the RKKY interaction, which often suppresses the Kondo screening and leads to the localization of $4f$ electrons. In addition, unconventional local-type quantum criticality has been proposed in some local-moment HF systems \cite{Si2001Locally,Coleman2001How,Custers2003nature,shishido2005JPSJ,Park2006CeRhIn5,Friedeman2010PNAS}, where the QCP is accompanied by an abrupt zero-temperature FS reconstruction, from a 'small FS' without $4f$ electrons in the AFM state to a 'large FS' including $4f$ electrons in the paramagnetic state. However, how the competition between the RKKY and Kondo interactions is manifested in the quasiparticle spectrum in a periodic Kondo lattice remains largely unexplored in electron spectroscopic experiments, e.g., angle-resolved photoemission spectroscopy (ARPES) and scanning tunneling microscopy (STM). Although it has been simulated by two-impurity Kondo systems in previous STM studies \cite{Bork2011tunable,Pruser2014interplay}, experiments on real periodic lattices are still lacking.

One approach to directly probe this competition is to track the $4f$ bands of an AFM HF system across (and well below) the magnetic transition using ARPES. However, due to the typically low transition temperatures in Ce-based HF systems, very few studies have been reported so far \cite{Poelchen2020unexpected}, and signatures of such competition have not been observed. CeCoGe$_{3}$ is a very suitable candidate for such a study, since it shows evidence of Kondo screening already at elevated temperatures and has a reasonably high AFM ordering temperature of \textit{T$_{N1}$} = 21 K \cite{Pecharsky1993PRB,Thamizhavel2005JPSJ,Michel2013neutronPRB}. Here we show that CeCoGe$_{3}$ does show signatures of the aforementioned competition, where our systematic temperature-dependent study across the transition reveals a clear deviation from Kondo-like behaviors in the AFM phase.

\begin{figure}[ht]
\includegraphics[width=1.\columnwidth]{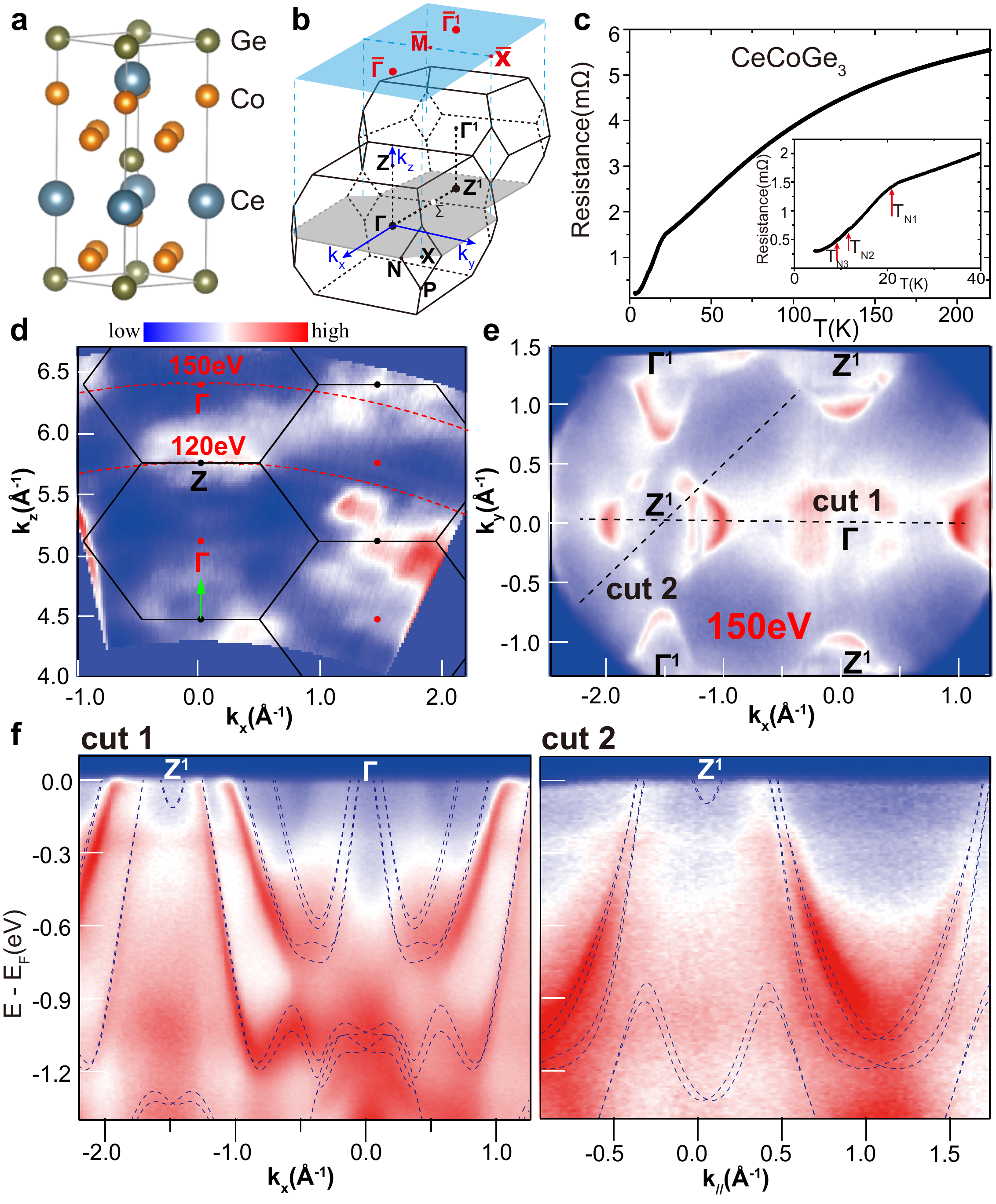}
\centering
\caption{(\textbf{a}) Crystal structure of CeCoGe$_{3}$. (\textbf{b}) The corresponding bulk and surface BZ's with high symmetry points labelled. (\textbf{c}) The resistance vs temperature for CeCoGe$_{3}$. The inset displays an enlarged view of the low temperature region showing the AFM transitions. (\textbf{d}) The $k_{x}-k_{z}$ FS map obtained from a photon-energy dependent scan (60 eV - 160 eV) at 26 K. The black hexagons indicate the bulk BZ boundaries and the dots indicate high symmetry momentum points. (\textbf{e}) In-plane $k_{x}-k_{y}$ FS map obtained from 150 eV photons ($k_{z}$ $\sim$ 0) at 5.2 K. (\textbf{f}) Energy-momentum dispersions along cut 1 and cut 2 shown in (\textbf{e}). Calculated bands (dashed blue curves) from DFT are overlaid on top.
}
\label{Fig1}
\end{figure}

CeCoGe$_3$ crystallizes in a noncentrosymmetric BaNiSn$_3$-type structure (Fig. \ref{Fig1}(a)), whose bulk Brillouin zone (BZ) and projected surface BZ are displayed in Fig. \ref{Fig1}(b). The non-centrosymmetric structure can give rise to Weyl points in its correlated band structure, as predicted by calculations \cite{Xiangangwan2021cal}, and unconventional superconductivity with mixed singlet-triplet pairing \cite{Settai2007JMMM,Frigeri2004}. The resistance (Fig. \ref{Fig1}(c)) clearly shows kinks at 21 K ($T_{N1}$), 12 K ($T_{N2}$) and 8 K ($T_{N3}$), which correspond to three consecutive AFM transitions \cite{Thamizhavel2005JPSJ}. Details of ARPES measurements and band structure calculations can be found in Refs. \cite{supplementary,Yang2021NST,PhysRevB.47.558,PhysRevB.59.1758,PhysRevLett.77.3865,MOSTOFI2008685}.


The $k_x-k_z$ and $k_x-k_y$ FS maps from ARPES measurements are summarized in Fig. \ref{Fig1}(d,e), which reveal diamond-shaped FS pockets centered at the bulk $\varGamma$ and \textit{Z} points (Fig. \ref{Fig1}(e)), respectively, with a periodicity in good accordance with the bulk BZ. A detailed photon-energy dependent study also shows well-defined $k_z$ periodicity of the observed bands (Fig. \ref{Fig1}(d)), which allows us to obtain an estimated inner potential of $\sim$15 eV. Hence, the 120 eV and 150 eV photon-energy scans correspond to $k_{z}$ $\sim$ $\pi$ and 0 (red dashed curves in Fig. \ref{Fig1}(d)), respectively. The energy-momentum dispersions are displayed in Fig. \ref{Fig1}(f), where the non-$4f$ conduction bands from density-functional theory (DFT) calculations are overlaid on top. The DFT calculations treat $4f$ electrons as core electrons and can provide a reasonable description of the observed conduction bands away from the Fermi level ($E_F$). However, the agreement near $E_F$ is not satisfactory, due to correlation effects from Ce $4f$ electrons (see below). We note that theoretical calculations predict a very small band splitting and Weyl points in the band structure as a result of the non-centrosymmetric crystal structure \cite{Xiangangwan2021cal}, but such splitting is too small to be resolved in our data.

A detailed comparison between the ARPES constant-energy maps and DFT calculations is shown in Fig. \ref{Fig2} (see Fig. S2 in \cite{supplementary} for more details). DFT calculations predict two hole pockets and one tiny electron pocket at \textit{Z}, as well as one large electron pocket and two small hole pockets at $\varGamma$, which partially agree with experiments, although some small pockets are absent experimentally [see also Fig. \ref{Fig1}(f)]. In general, the constant energy maps away from $E_F$ show better agreement with DFT calculations, compared to those close to $E_F$, reflecting the strong local correlations of the $4f$ electrons, which usually affects the quasiparticle dispersion only in the vicinity of $E_F$. A previous de Haas-van Alphen (dHvA) effect study  at very low temperatures indicates that the FS shape of CeCoGe$_3$ is overall similar to that of LaCoGe$_3$ with only small differences \cite{Thamizhavel2006JPSJ}, while the cyclotron mass of the dominant dHvA branch is ten times higher for CeCoGe$_3$ ($\sim$10 $m_e$ for CeCoGe$_3$ vs $\sim$1 $m_e$ for LaCoGe$_3$). This implies that although the overall FS shape of CeCoGe$_3$ is not significantly altered by the $4f$ electrons, appreciable contributions from $4f$ electrons are present in the FS due to the Kondo effect.

\begin{figure}[ht]
\includegraphics[width=0.9\columnwidth]{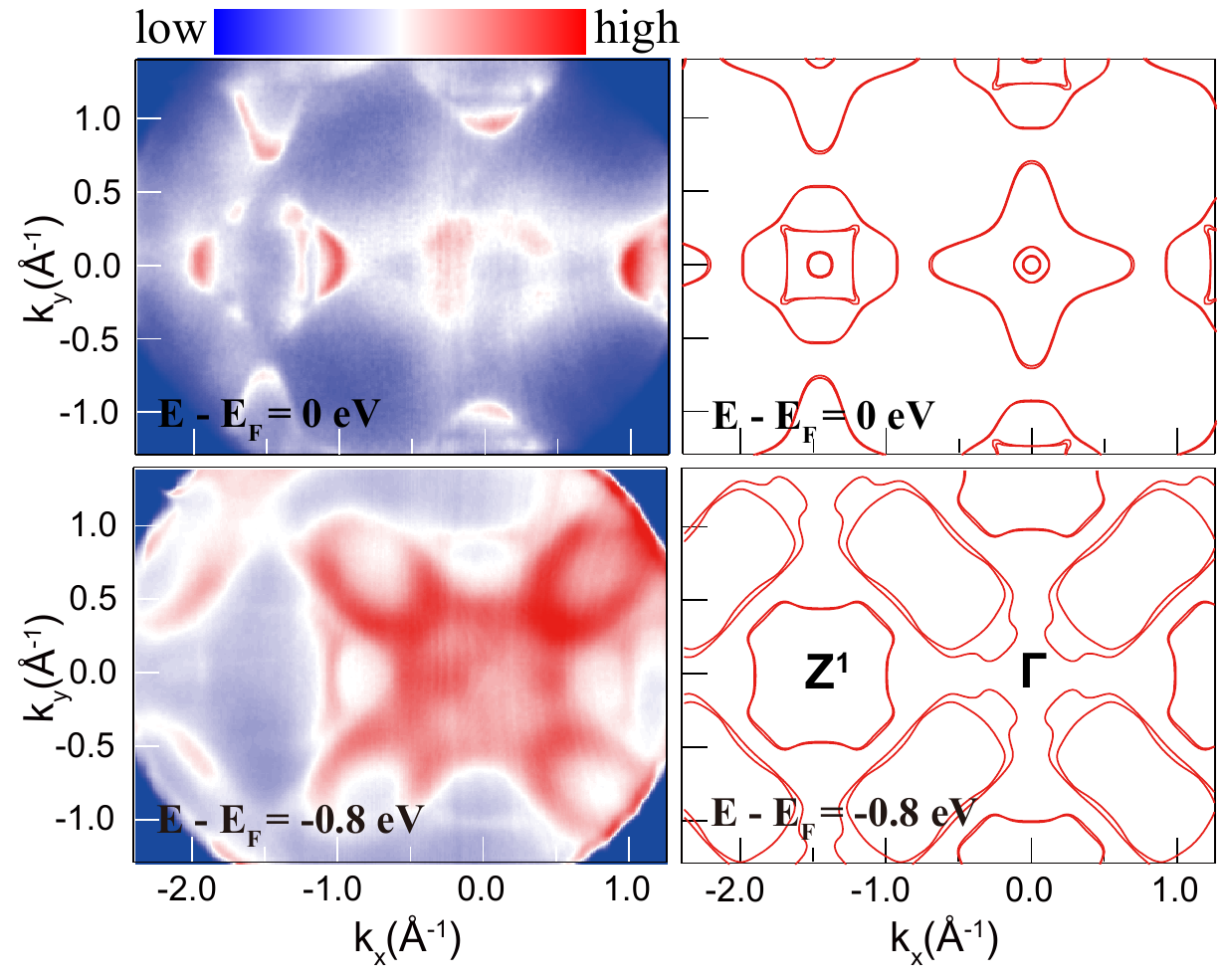}
\centering
\caption{Left: constant energy contours at $E$ = 0 and -0.8 eV using 150 eV photons at 5.2 K. Right: the corresponding DFT calculations.
}
\label{Fig2}
\end{figure}

\begin{figure}[ht]
\includegraphics[width=1.\columnwidth]{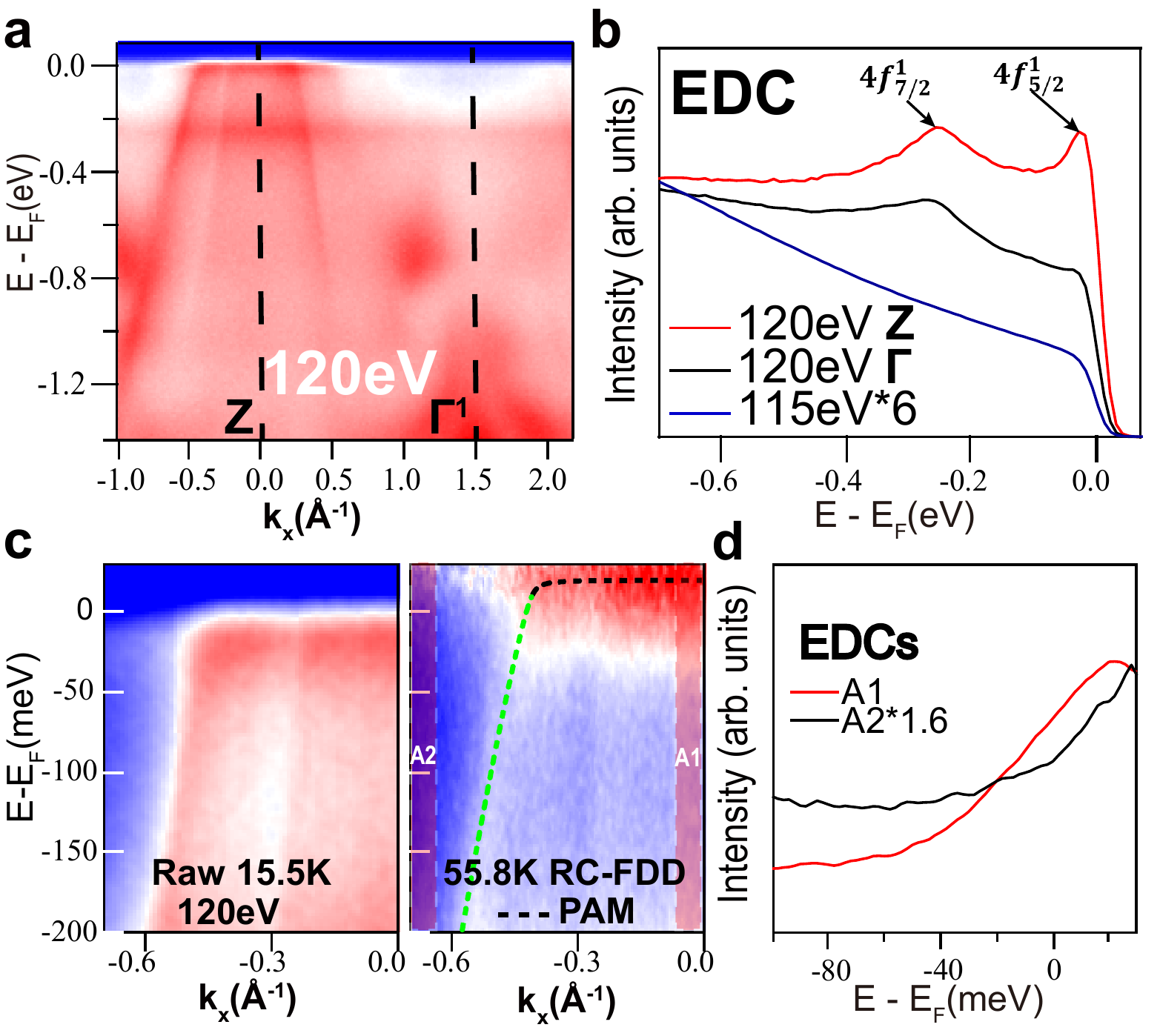}
\centering
\caption{(\textbf{a}) Band dispersion along $\varGamma$-\textit{Z} from resonant ARPES. (\textbf{b}) Resonant EDCs from the black dashed cuts in (\textbf{a}), together with the off-resonant EDC at 115 eV. (\textbf{c}) Left: zoom-in view of the quasiparticle dispersion at 15.5 K near $E_F$. Right: the 56 K data divided by the RC-FDD. The dotted curve is the fitted band dispersion using Eq. (1). (\textbf{d}) Two integrated EDCs from the RC-FDD-divided 56 K data, for two momentum regions (A1 and A2) marked in (\textbf{c}).
}
\label{Fig3}
\end{figure}

To probe the $4f$ states, we performed resonant ARPES measurements at the Ce $N$ edge near 120 eV, which significantly enhances the $4f$ contribution to the measured spectra. The resonant ARPES spectrum (Fig. \ref{Fig3}(a)) features a flat $4f$ band right at $E_F$ near the \textit{Z} point. This is better illustrated by the energy distribution curves (EDCs) shown in Fig. \ref{Fig3}(b), where clear resonance enhancement of the peaks at $E_F$ and -0.25 eV can be observed, corresponding to the $4f^1_{5/2}$ and $4f^1_{7/2}$ peaks, respectively \cite{jw2005kondo,sekiyama2000probing}. Although the $4f^1_{7/2}$ peak can be clearly observed at all momentum points, the $4f^1_{5/2}$ peak exhibits a large momentum dependence (see Fig. S3 in \cite{supplementary} for detailed FS maps).
In particular, the absence of a $4f^1_{5/2}$ peak around the $\varGamma$ point may be attributed to a small indirect gap as a result of hybridization between conduction and $4f$ electrons ($c-f$ hybridization). The $c-f$ hybridization can also give rise to weak band bending near $E_F$ \cite{Im2008,patil2016arpes,Chen2017,Jang2020,WuCeRh6Ge4PRL2021,kirchner2020colloquium}, as shown in Fig. \ref{Fig3}(c). To reveal the fine dispersion near $E_F$, we divide the ARPES spectra at 56 K by the resolution-convoluted Fermi-Dirac distribution (RC-FDD), obtained by fitting the reference spectrum. The result (right panel in Fig. \ref{Fig3}(c)) reveals a weakly dispersive $4f$ band whose center lies slightly above $E_F$. For quantitative analysis, we adopted the hybridized band approach within the periodic Anderson model (PAM), where the band dispersion can be described by
\begin{equation}\label{1}
  E^\pm(k)=\frac{\overline{\varepsilon}_f+\varepsilon_k\pm\sqrt{(\overline{\varepsilon}_f-\varepsilon_k)^2+4V^2}}{2}.
\end{equation}
Here, $\overline{\varepsilon}_f$ and $\varepsilon_k$ are the energies of the single-ion $4f_{5/2}^1$ Kondo resonance and the conduction band in the absence of $c-f$ hybridization, respectively, and $V$ is the effective $c-f$ band hybridization strength.
Our analysis is focused on the lower-energy branch in Eq. (1), which crosses $E_F$, and we assume that $\varepsilon_k$ is linear in $k$ in the vicinity of $E_F$. The peak positions can be obtained by analyzing the momentum distribution curves (MDCs) and EDCs (Fig. S4 in \cite{supplementary}), which can then be fitted by Eq. (1) to obtain $V$ $\sim$ 10 meV, indicating a moderate $c-f$ hybridization. The RC-FDD divided EDCs in Fig. \ref{Fig3}(d) further reveal clear momentum dependence of the positions of the $4f$ bands (hence an indirect gap caused by the $c-f$ hybridization), which could explain the very weak $4f^1_{5/2}$ peak far away from \textit{Z}.

\begin{figure}[ht]
\includegraphics[width=1.\columnwidth]{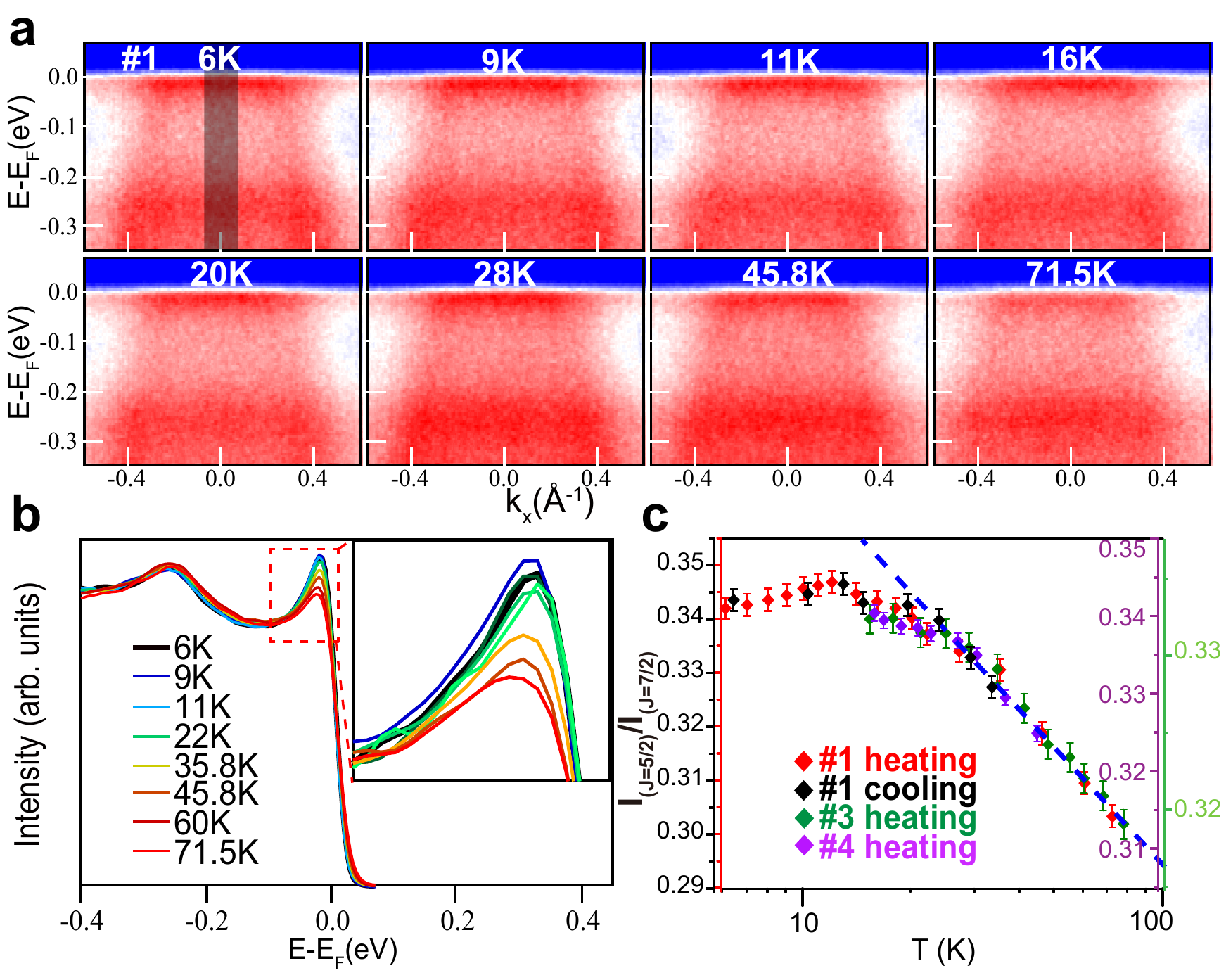}
\centering
\caption{(\textbf{a}) Resonant ARPES spectra near \textit{Z} at various temperatures for a typical sample (\#1). (\textbf{b}) The temperature-dependent EDCs obtained by integrating the shaded region in (\textbf{a}). Inset is an enlarged view of the $4f^1_{5/2}$ peak. (\textbf{c}) The ratio between the integrated intensity of $4f^1_{5/2}$ ([-0.09 eV, 0 eV]) and that of $4f^1_{7/2}$ ([-0.38 eV, -0.12 eV]), as a function of temperature. The blue dashed curve indicates the $-\log(T)$ reference line. Note the slightly different vertical scales due to sample variation.
}
\label{Fig4}
\end{figure}

Temperature-dependent ARPES measurements near the \textit{Z} point are summarized in Fig. \ref{Fig4}(a,b). We first note that the $4f^1_{5/2}$ peak can already be observed at $\sim$80 K (and above), similar to other HF systems \cite{Chen2017,Jang2020,ehm2007high}, implying that Kondo screening takes place already at high temperatures, presumably involving excited crystal field states. Upon cooling, the $4f^1_{5/2}$ peak becomes more prominent due to the gradual buildup of heavy quasiparticles from the Kondo effect, while the $4f^1_{7/2}$ peak at -0.25 eV remains largely unchanged, probably due to its much higher state degeneracy (hence higher $T_K$) \cite{Reinert2001,Kroha2003}. In Fig. \ref{Fig4}(c), we plot the temperature evolution of the relative intensity of the $4f^1_{5/2}$ peak, i.e., the ratio between the integrated intensity of $4f^1_{5/2}$ and that of $4f^1_{7/2}$, which reduces extrinsic effects that might affect the normalization of the temperature-dependent data (see Fig. S5 in \cite{supplementary} for details). The relative $4f^1_{5/2}$ intensity roughly follows a $-\log(T)$ behavior from $\sim$25 K and above, as expected for a Kondo peak in the paramagnetic state \cite{Chen2017,Jang2020}, but it increases more slowly upon cooling below $T_1\sim$25 K and eventually ceases to grow below $T_2\sim$12~K. We emphasize that such a temperature evolution has been confirmed by measurements from different samples (Fig. S6 in \cite{supplementary}), as well as in different cooling and warming cycles from the same sample. The results indicate that the heavy quasiparticle formation due to the Kondo effect begins to slow down below $T_1$, and stops (or becomes slightly suppressed) below $T_2$. The fact that $T_1$ is close to $T_{N1}=21$ K implies that the weakening of the Kondo processes is very likely caused by the development of long-range AFM order. Our results therefore show that the magnetic order and Kondo screening coexist and compete over a large temperature window, even deep inside the AFM phase.

Our observation is in stark contrast to those from previous ARPES measurements of the Ce-based AFM Kondo systems CeRh$_2$Si$_2$ and CePt$_2$In$_7$. For the former with a very high $T_N=42$~K, the Kondo peak from ARPES measurements shows a monotonic increase with cooling, without any noticeable deviation even well below $T_N$ \cite{Poelchen2020unexpected}. For CePt$_2$In$_7$, the suppression of a weak Kondo peak occurs below $\sim$60 K \cite{Luo2020PRB}, much higher than $T_N=5$~K, which, however, is different from the results inferred from Knight shift measurements which imply a relocalization of the $4f$ electrons only below $\sim$12 K \cite{apRoberts2011PRB}. 
We note that recent ARPES studies on the 5\textit{f}-electron compounds USb$_{2}$ and UAs$_{2}$ show that the magnetic order and Kondo effect could coexist without interfering with each other \cite{Chen2019Orbital,Ji2022PRB}. Specifically, ARPES measurements revealed two sets of $5f$-derived quasiparticle bands that are well separated in momentum space \cite{Chen2019Orbital,Ji2022PRB}: in one band, the itinerant SDW order opens up a gap (hence accounting for the magnetism), while the other band corresponds to the typical heavy quasiparticles found in HF systems where the peak intensity remains unaffected by the development of AFM order. The differences between the U-based and Ce-based compounds could be attributed to the different number of $f$ electrons: in U-based compounds, U normally has three electrons in the $5f$ orbitals which may selectively participate in the Kondo screening or AFM order; by contrast, in Ce there is only one electron in the $4f$ shell, so that such orbital selectivity is not possible. Therefore, in Ce-based HF materials, the competition and coexistence of Kondo effect and AFM order must be of fundamentally different nature.

The origin of this competing behavior in CeCoGe$_3$ may lie in the localized nature of the 4$f$-electron magnetism, namely, the AFM order consists of Ising-like moments pointing along $z$ that are coupled ferromagnetically in the $x-y$ plane and antiferromagnetically along the $z$ direction with different ordering vectors $\mathbf{k}=(0,0,q_z)$ in the three magnetic phases \cite{Michel2013neutronPRB}. The commensurate AFM orderings in CeCoGe$_3$ are unlikely to correspond to the itinerant SDW type, and the observed FS does not show any sign of nesting. In addition, temperature-dependent FS mappings show no obvious change in the FS shape across the AFM transitions (Fig. S7 in \cite{supplementary}), supporting the local-moment nature of the AFM order.

The observation of coexistence and competition between the magnetic order and the Kondo effect in Fig. \ref{Fig4}(c) poses the question as to the nature of the microscopic interplay between these phenomena. One possible scenario is that both the ordered magnetic moments and Kondo screening are modulated in real space, such that there is a periodic arrangement of sizeable ordered moments and Kondo-screened moments. Such a picture was also proposed based on a previous neutron scattering study \cite{Broholm2019}, while an analogous proposal was put forward for UAu$_{2}$ based on a two-channel Kondo model for multiple $5f$ electrons \cite{Neill2021}. Such a scenario could account for the significant Kondo weight persisting well below \textit{T$_{N}$}, while the close interplay between the magnetic structure and Kondo screening may also explain why the turning point \textit{T$_{2}$} is close to \textit{T$_{N2}$} (and \textit{T$_{N3}$}), where there is a change of magnetic structure \cite{Michel2013neutronPRB}.

Interestingly, for the experimental $k_x-k_z$ map in Fig. \ref{Fig1}(d), the dominant pocket observed near \textit{Z} has a Fermi vector $k_F$ (marked by a green arrow in Fig. \ref{Fig1}(d)) that is approximately 1/2 of $\varGamma$-\textit{Z}. Therefore, the oscillatory RKKY interaction along $z$ could have a node for interactions between nearest-neighbor Ce planes (note that for the body-centered structure there are two Ce atoms per conventional unit cell separated by $c$/2 along $z$ [Fig. \ref{Fig1}(a)]). Below $T_{N3}$, where $q_z=1/2$, the AFM order could lead to adjacent Ce layers alternating between large ordered moments (antiparallel between
next-nearest layers) and nearly zero moments (Kondo screened) (Fig. S8 in \cite{supplementary}). Note that although an equal moment ``two-up two-down'' structure was previously proposed \cite{Michel2013neutronPRB}, the neutron diffraction results are equally compatible with the modulated structure described here, after a shift of the global phase by $\pi/4$.

An alternative scenario would be that there is an inherent onsite competition and coexistence of magnetic order and Kondo screening. In this case, the Kondo weight is expected to decrease inside the AFM phase, or even reach zero in the zero-temperature limit if the AFM QCP is of the local type \cite{Coleman2007heavy,gegenwart2008quantum,Kroha2017Interplay,kirchner2020colloquium,Yang2012PNAS}. Although the current results suggest a very small decrease of Kondo weight below $T_2$ (Fig. \ref{Fig4}(c)), the weight appears to remain finite based upon a simple extrapolation to lower temperatures. On the other hand, since the current measurements are performed only down to 6~K, determining the fate of the Kondo peak in the zero-temperature limit requires future measurements to significantly lower temperatures.

To conclude, using high-resolution ARPES, we probed the quasiparticle bands near $E_F$ in CeCoGe$_{3}$ with a high $T_{N1}=21$~K. The presence of a well-resolved heavy $4f$ band with a relatively simple momentum distribution allows us to clearly track the temperature evolution of the $4f$ band across the AFM transitions, and we find that its intensity starts to deviate from the typical $-\log(T)$ behavior just above $T_{\rm N1}$, and eventually ceases to grow at lower temperatures. Our results provide direct spectroscopic evidence for the competition between the well-localized magnetic order and Kondo effect in a HF system, which might be related to a real-space modulation of magnetic moments and Kondo screening. These findings may be useful for understanding other HF systems with local-moment antiferromagnetism, e.g., CeRhIn$_5$, where a 'small FS' with enhanced cyclotron masses is observed below $T_{N}$ from dHvA measurements \cite{shishido2002JPSJ,Harrison2004PRL}, while ARPES reveals weakly dispersive $4f$-bands well above $T_{N}$ from moderate $c-f$ hybridization \cite{Chen2018}, similar to CeCoGe$_{3}$. Nevertheless, the low ordering temperature of $T_{\rm N}=3.8$~K in CeRhIn$_5$ has precluded direct observation of such competition. Finally, since the degree of $f$-electron itinerancy and the nature of the magnetic ground state can be readily tuned by elemental substitution in the Ce$TX_3$ series \cite{Thamizhavel2006JPSJ,Okuda2007,Terashima2008,Ohkochi2009JPSJ,Smidman2015PRBCeRuSi3,pasztorova2015}, these systems could be ideal candidates for systematically studying the electronic manifestation of the competition between the RKKY and Kondo interactions.


~
\par This work is supported by the Key R$\&$D Program of Zhejiang Province, China (2021C01002), NSF of China (No. 12174331, 12034017, U2032208, 12174332, 11974397), National Key R$\&$D Program of the MOST of China (Grant No. 2022YFA1402200, 2017YFA0303100), the State Key project of Zhejiang Province (No. Z22A049846), the Fundamental Research Funds for the Central Universities (2021FZZX001-03) and Bridging Grants with China, Japan, South Korea and ASEAN region (BG11-072020) funded by the Swiss State Secretariat. Part of this research used Beamline 03U at SSRF, supported by ME2 project under Contract No. 11227902 from NSF of China. The work at the University of Warwick was supported by EPSRC, UK, through Grants EP/M028771/1 and EP/T005963/1. J.K. was supported in part by the Deutsche Forschungsgemeinschaft through the Cluster of Excellence ML4Q (90534769).

\bibliography{CCGv12}

\begin{thebibliography}{68}%
\makeatletter
\providecommand \@ifxundefined [1]{%
 \@ifx{#1\undefined}
}%
\providecommand \@ifnum [1]{%
 \ifnum #1\expandafter \@firstoftwo
 \else \expandafter \@secondoftwo
 \fi
}%
\providecommand \@ifx [1]{%
 \ifx #1\expandafter \@firstoftwo
 \else \expandafter \@secondoftwo
 \fi
}%
\providecommand \natexlab [1]{#1}%
\providecommand \enquote  [1]{``#1''}%
\providecommand \bibnamefont  [1]{#1}%
\providecommand \bibfnamefont [1]{#1}%
\providecommand \citenamefont [1]{#1}%
\providecommand \href@noop [0]{\@secondoftwo}%
\providecommand \href [0]{\begingroup \@sanitize@url \@href}%
\providecommand \@href[1]{\@@startlink{#1}\@@href}%
\providecommand \@@href[1]{\endgroup#1\@@endlink}%
\providecommand \@sanitize@url [0]{\catcode `\\12\catcode `\$12\catcode
  `\&12\catcode `\#12\catcode `\^12\catcode `\_12\catcode `\%12\relax}%
\providecommand \@@startlink[1]{}%
\providecommand \@@endlink[0]{}%
\providecommand \url  [0]{\begingroup\@sanitize@url \@url }%
\providecommand \@url [1]{\endgroup\@href {#1}{\urlprefix }}%
\providecommand \urlprefix  [0]{URL }%
\providecommand \Eprint [0]{\href }%
\providecommand \doibase [0]{https://doi.org/}%
\providecommand \selectlanguage [0]{\@gobble}%
\providecommand \bibinfo  [0]{\@secondoftwo}%
\providecommand \bibfield  [0]{\@secondoftwo}%
\providecommand \translation [1]{[#1]}%
\providecommand \BibitemOpen [0]{}%
\providecommand \bibitemStop [0]{}%
\providecommand \bibitemNoStop [0]{.\EOS\space}%
\providecommand \EOS [0]{\spacefactor3000\relax}%
\providecommand \BibitemShut  [1]{\csname bibitem#1\endcsname}%
\let\auto@bib@innerbib\@empty
\bibitem [{\citenamefont {Coleman}(2007)}]{Coleman2007heavy}%
  \BibitemOpen
  \bibfield  {author} {\bibinfo {author} {\bibfnamefont {P.}~\bibnamefont
  {Coleman}},\ }\href@noop {} {\emph {\bibinfo {title} {Heavy Fermions:
  electrons at the edge of magnetism. Handbook of Magnetism and Advanced
  Magnetic Materials}}},\ Vol.~\bibinfo {volume} {1}\ (\bibinfo  {publisher}
  {Wiley, New York},\ \bibinfo {year} {2007})\BibitemShut {NoStop}%
\bibitem [{\citenamefont {Gegenwart}\ \emph {et~al.}(2008)\citenamefont
  {Gegenwart}, \citenamefont {Si},\ and\ \citenamefont
  {Steglich}}]{gegenwart2008quantum}%
  \BibitemOpen
  \bibfield  {author} {\bibinfo {author} {\bibfnamefont {P.}~\bibnamefont
  {Gegenwart}}, \bibinfo {author} {\bibfnamefont {Q.}~\bibnamefont {Si}},\ and\
  \bibinfo {author} {\bibfnamefont {F.}~\bibnamefont {Steglich}},\ }\href
  {https://doi.org/10.1038/nphys892} {\bibfield  {journal} {\bibinfo  {journal}
  {Nature physics}\ }\textbf {\bibinfo {volume} {4}},\ \bibinfo {pages} {186}
  (\bibinfo {year} {2008})}\BibitemShut {NoStop}%
\bibitem [{\citenamefont {L\"ohneysen}\ \emph {et~al.}(2007)\citenamefont
  {L\"ohneysen}, \citenamefont {Rosch}, \citenamefont {Vojta},\ and\
  \citenamefont {W\"olfle}}]{Lohneysen2007Fermi}%
  \BibitemOpen
  \bibfield  {author} {\bibinfo {author} {\bibfnamefont {H.~v.}\ \bibnamefont
  {L\"ohneysen}}, \bibinfo {author} {\bibfnamefont {A.}~\bibnamefont {Rosch}},
  \bibinfo {author} {\bibfnamefont {M.}~\bibnamefont {Vojta}},\ and\ \bibinfo
  {author} {\bibfnamefont {P.}~\bibnamefont {W\"olfle}},\ }\href
  {https://doi.org/10.1103/RevModPhys.79.1015} {\bibfield  {journal} {\bibinfo
  {journal} {Rev. Mod. Phys.}\ }\textbf {\bibinfo {volume} {79}},\ \bibinfo
  {pages} {1015} (\bibinfo {year} {2007})}\BibitemShut {NoStop}%
\bibitem [{\citenamefont {Sachdev}(2011)}]{Sachdev2011Quantum}%
  \BibitemOpen
  \bibfield  {author} {\bibinfo {author} {\bibfnamefont {S.}~\bibnamefont
  {Sachdev}},\ }\href@noop {} {\emph {\bibinfo {title} {Quantum Phase
  Transitions}}}\ (\bibinfo  {publisher} {Cambridge Univ. Press},\ \bibinfo
  {year} {2011})\BibitemShut {NoStop}%
\bibitem [{\citenamefont {Steglich}\ \emph {et~al.}(1979)\citenamefont
  {Steglich}, \citenamefont {Aarts}, \citenamefont {Bredl}, \citenamefont
  {Lieke}, \citenamefont {Meschede}, \citenamefont {Franz},\ and\ \citenamefont
  {Sch\"afer}}]{Steglich1979Superconductivity}%
  \BibitemOpen
  \bibfield  {author} {\bibinfo {author} {\bibfnamefont {F.}~\bibnamefont
  {Steglich}}, \bibinfo {author} {\bibfnamefont {J.}~\bibnamefont {Aarts}},
  \bibinfo {author} {\bibfnamefont {C.~D.}\ \bibnamefont {Bredl}}, \bibinfo
  {author} {\bibfnamefont {W.}~\bibnamefont {Lieke}}, \bibinfo {author}
  {\bibfnamefont {D.}~\bibnamefont {Meschede}}, \bibinfo {author}
  {\bibfnamefont {W.}~\bibnamefont {Franz}},\ and\ \bibinfo {author}
  {\bibfnamefont {H.}~\bibnamefont {Sch\"afer}},\ }\href
  {https://doi.org/10.1103/PhysRevLett.43.1892} {\bibfield  {journal} {\bibinfo
   {journal} {Phys. Rev. Lett.}\ }\textbf {\bibinfo {volume} {43}},\ \bibinfo
  {pages} {1892} (\bibinfo {year} {1979})}\BibitemShut {NoStop}%
\bibitem [{\citenamefont {Mathur}\ \emph {et~al.}(1998)\citenamefont {Mathur},
  \citenamefont {Grosche}, \citenamefont {Julian}, \citenamefont {Walker},
  \citenamefont {Freye}, \citenamefont {Haselwimmer},\ and\ \citenamefont
  {Lonzarich}}]{Mathur1998Magnetically}%
  \BibitemOpen
  \bibfield  {author} {\bibinfo {author} {\bibfnamefont {N.~D.}\ \bibnamefont
  {Mathur}}, \bibinfo {author} {\bibfnamefont {F.~M.}\ \bibnamefont {Grosche}},
  \bibinfo {author} {\bibfnamefont {S.~R.}\ \bibnamefont {Julian}}, \bibinfo
  {author} {\bibfnamefont {I.~R.}\ \bibnamefont {Walker}}, \bibinfo {author}
  {\bibfnamefont {D.~M.}\ \bibnamefont {Freye}}, \bibinfo {author}
  {\bibfnamefont {R.~K.~W.}\ \bibnamefont {Haselwimmer}},\ and\ \bibinfo
  {author} {\bibfnamefont {G.~G.}\ \bibnamefont {Lonzarich}},\ }\href
  {https://doi.org/10.1038/27838} {\bibfield  {journal} {\bibinfo  {journal}
  {Nature}\ }\textbf {\bibinfo {volume} {394}},\ \bibinfo {pages} {39}
  (\bibinfo {year} {1998})}\BibitemShut {NoStop}%
\bibitem [{\citenamefont {Stewart}(2001)}]{Stewart2001NFL}%
  \BibitemOpen
  \bibfield  {author} {\bibinfo {author} {\bibfnamefont {G.~R.}\ \bibnamefont
  {Stewart}},\ }\href {https://doi.org/10.1103/RevModPhys.73.797} {\bibfield
  {journal} {\bibinfo  {journal} {Rev. Mod. Phys.}\ }\textbf {\bibinfo {volume}
  {73}},\ \bibinfo {pages} {797} (\bibinfo {year} {2001})}\BibitemShut
  {NoStop}%
\bibitem [{\citenamefont {Ruderman}\ and\ \citenamefont {Kittel}(1954)}]{RK}%
  \BibitemOpen
  \bibfield  {author} {\bibinfo {author} {\bibfnamefont {M.~A.}\ \bibnamefont
  {Ruderman}}\ and\ \bibinfo {author} {\bibfnamefont {C.}~\bibnamefont
  {Kittel}},\ }\href {https://doi.org/10.1103/PhysRev.96.99} {\bibfield
  {journal} {\bibinfo  {journal} {Phys. Rev.}\ }\textbf {\bibinfo {volume}
  {96}},\ \bibinfo {pages} {99} (\bibinfo {year} {1954})}\BibitemShut {NoStop}%
\bibitem [{\citenamefont {Kasuya}(1956)}]{K}%
  \BibitemOpen
  \bibfield  {author} {\bibinfo {author} {\bibfnamefont {T.}~\bibnamefont
  {Kasuya}},\ }\href {https://doi.org/10.1143/PTP.16.45} {\bibfield  {journal}
  {\bibinfo  {journal} {Prog. Theor. Phys.}\ }\textbf {\bibinfo {volume}
  {16}},\ \bibinfo {pages} {45} (\bibinfo {year} {1956})}\BibitemShut {NoStop}%
\bibitem [{\citenamefont {Yosida}(1957)}]{Y}%
  \BibitemOpen
  \bibfield  {author} {\bibinfo {author} {\bibfnamefont {K.}~\bibnamefont
  {Yosida}},\ }\href {https://doi.org/10.1103/PhysRev.106.893} {\bibfield
  {journal} {\bibinfo  {journal} {Phys. Rev.}\ }\textbf {\bibinfo {volume}
  {106}},\ \bibinfo {pages} {893} (\bibinfo {year} {1957})}\BibitemShut
  {NoStop}%
\bibitem [{\citenamefont {Kondo}(1964)}]{Kondo1964}%
  \BibitemOpen
  \bibfield  {author} {\bibinfo {author} {\bibfnamefont {J.}~\bibnamefont
  {Kondo}},\ }\href {https://doi.org/10.1143/PTP.32.37} {\bibfield  {journal}
  {\bibinfo  {journal} {Prog. Theor. Phys.}\ }\textbf {\bibinfo {volume}
  {32}},\ \bibinfo {pages} {37} (\bibinfo {year} {1964})}\BibitemShut {NoStop}%
\bibitem [{\citenamefont {Hewson}(1993)}]{Hewson1993Kondo}%
  \BibitemOpen
  \bibfield  {author} {\bibinfo {author} {\bibfnamefont {A.~C.}\ \bibnamefont
  {Hewson}},\ }\href@noop {} {\emph {\bibinfo {title} {The Kondo Problem to
  Heavy Fermions}}}\ (\bibinfo  {publisher} {Cambridge Univ. Press},\ \bibinfo
  {year} {1993})\BibitemShut {NoStop}%
\bibitem [{\citenamefont {Nejati}\ \emph {et~al.}(2017)\citenamefont {Nejati},
  \citenamefont {Ballmann},\ and\ \citenamefont {Kroha}}]{Nejati2017}%
  \BibitemOpen
  \bibfield  {author} {\bibinfo {author} {\bibfnamefont {A.}~\bibnamefont
  {Nejati}}, \bibinfo {author} {\bibfnamefont {K.}~\bibnamefont {Ballmann}},\
  and\ \bibinfo {author} {\bibfnamefont {J.}~\bibnamefont {Kroha}},\ }\href
  {https://doi.org/10.1103/PhysRevLett.118.117204} {\bibfield  {journal}
  {\bibinfo  {journal} {Phys. Rev. Lett.}\ }\textbf {\bibinfo {volume} {118}},\
  \bibinfo {pages} {117204} (\bibinfo {year} {2017})}\BibitemShut {NoStop}%
\bibitem [{\citenamefont {Kirchner}\ \emph {et~al.}(2020)\citenamefont
  {Kirchner}, \citenamefont {Paschen}, \citenamefont {Chen}, \citenamefont
  {Wirth}, \citenamefont {Feng}, \citenamefont {Thompson},\ and\ \citenamefont
  {Si}}]{kirchner2020colloquium}%
  \BibitemOpen
  \bibfield  {author} {\bibinfo {author} {\bibfnamefont {S.}~\bibnamefont
  {Kirchner}}, \bibinfo {author} {\bibfnamefont {S.}~\bibnamefont {Paschen}},
  \bibinfo {author} {\bibfnamefont {Q.}~\bibnamefont {Chen}}, \bibinfo {author}
  {\bibfnamefont {S.}~\bibnamefont {Wirth}}, \bibinfo {author} {\bibfnamefont
  {D.}~\bibnamefont {Feng}}, \bibinfo {author} {\bibfnamefont {J.~D.}\
  \bibnamefont {Thompson}},\ and\ \bibinfo {author} {\bibfnamefont
  {Q.}~\bibnamefont {Si}},\ }\href
  {https://doi.org/10.1103/RevModPhys.92.011002} {\bibfield  {journal}
  {\bibinfo  {journal} {Rev. Mod. Phys.}\ }\textbf {\bibinfo {volume} {92}},\
  \bibinfo {pages} {011002} (\bibinfo {year} {2020})}\BibitemShut {NoStop}%
\bibitem [{\citenamefont {Yang}\ and\ \citenamefont
  {Pines}(2012)}]{Yang2012PNAS}%
  \BibitemOpen
  \bibfield  {author} {\bibinfo {author} {\bibfnamefont {Y.-f.}\ \bibnamefont
  {Yang}}\ and\ \bibinfo {author} {\bibfnamefont {D.}~\bibnamefont {Pines}},\
  }\href {https://doi.org/10.1073/pnas.1211186109} {\bibfield  {journal}
  {\bibinfo  {journal} {Proc. Natl. Acad. Sci.}\ }\textbf {\bibinfo {volume}
  {109}},\ \bibinfo {pages} {E3060} (\bibinfo {year} {2012})}\BibitemShut
  {NoStop}%
\bibitem [{\citenamefont {Doniach}(1977)}]{Doniach1977Kondo}%
  \BibitemOpen
  \bibfield  {author} {\bibinfo {author} {\bibfnamefont {S.}~\bibnamefont
  {Doniach}},\ }\href
  {https://doi.org/https://doi.org/10.1016/0378-4363(77)90190-5} {\bibfield
  {journal} {\bibinfo  {journal} {Physica B \& C}\ }\textbf {\bibinfo {volume}
  {91}},\ \bibinfo {pages} {231} (\bibinfo {year} {1977})}\BibitemShut
  {NoStop}%
\bibitem [{\citenamefont {Zwicknagl}\ and\ \citenamefont
  {Pulst}(1993)}]{Zwicknagl1993Physica}%
  \BibitemOpen
  \bibfield  {author} {\bibinfo {author} {\bibfnamefont {G.}~\bibnamefont
  {Zwicknagl}}\ and\ \bibinfo {author} {\bibfnamefont {U.}~\bibnamefont
  {Pulst}},\ }\href
  {https://doi.org/https://doi.org/10.1016/0921-4526(93)90736-P} {\bibfield
  {journal} {\bibinfo  {journal} {Physica B}\ }\textbf {\bibinfo {volume}
  {186-188}},\ \bibinfo {pages} {895} (\bibinfo {year} {1993})}\BibitemShut
  {NoStop}%
\bibitem [{\citenamefont {Gegenwart}\ \emph {et~al.}(1998)\citenamefont
  {Gegenwart}, \citenamefont {Langhammer}, \citenamefont {Geibel},
  \citenamefont {Helfrich}, \citenamefont {Lang}, \citenamefont {Sparn},
  \citenamefont {Steglich}, \citenamefont {Horn}, \citenamefont {Donnevert},
  \citenamefont {Link},\ and\ \citenamefont {Assmus}}]{Gegenwart1998}%
  \BibitemOpen
  \bibfield  {author} {\bibinfo {author} {\bibfnamefont {P.}~\bibnamefont
  {Gegenwart}}, \bibinfo {author} {\bibfnamefont {C.}~\bibnamefont
  {Langhammer}}, \bibinfo {author} {\bibfnamefont {C.}~\bibnamefont {Geibel}},
  \bibinfo {author} {\bibfnamefont {R.}~\bibnamefont {Helfrich}}, \bibinfo
  {author} {\bibfnamefont {M.}~\bibnamefont {Lang}}, \bibinfo {author}
  {\bibfnamefont {G.}~\bibnamefont {Sparn}}, \bibinfo {author} {\bibfnamefont
  {F.}~\bibnamefont {Steglich}}, \bibinfo {author} {\bibfnamefont
  {R.}~\bibnamefont {Horn}}, \bibinfo {author} {\bibfnamefont {L.}~\bibnamefont
  {Donnevert}}, \bibinfo {author} {\bibfnamefont {A.}~\bibnamefont {Link}},\
  and\ \bibinfo {author} {\bibfnamefont {W.}~\bibnamefont {Assmus}},\ }\href
  {https://doi.org/10.1103/PhysRevLett.81.1501} {\bibfield  {journal} {\bibinfo
   {journal} {Phys. Rev. Lett.}\ }\textbf {\bibinfo {volume} {81}},\ \bibinfo
  {pages} {1501} (\bibinfo {year} {1998})}\BibitemShut {NoStop}%
\bibitem [{\citenamefont {Stockert}\ \emph {et~al.}(2004)\citenamefont
  {Stockert}, \citenamefont {Faulhaber}, \citenamefont {Zwicknagl},
  \citenamefont {St\"u\ss{}er}, \citenamefont {Jeevan}, \citenamefont {Deppe},
  \citenamefont {Borth}, \citenamefont {K\"uchler}, \citenamefont
  {Loewenhaupt}, \citenamefont {Geibel},\ and\ \citenamefont
  {Steglich}}]{Stockert2004}%
  \BibitemOpen
  \bibfield  {author} {\bibinfo {author} {\bibfnamefont {O.}~\bibnamefont
  {Stockert}}, \bibinfo {author} {\bibfnamefont {E.}~\bibnamefont {Faulhaber}},
  \bibinfo {author} {\bibfnamefont {G.}~\bibnamefont {Zwicknagl}}, \bibinfo
  {author} {\bibfnamefont {N.}~\bibnamefont {St\"u\ss{}er}}, \bibinfo {author}
  {\bibfnamefont {H.~S.}\ \bibnamefont {Jeevan}}, \bibinfo {author}
  {\bibfnamefont {M.}~\bibnamefont {Deppe}}, \bibinfo {author} {\bibfnamefont
  {R.}~\bibnamefont {Borth}}, \bibinfo {author} {\bibfnamefont
  {R.}~\bibnamefont {K\"uchler}}, \bibinfo {author} {\bibfnamefont
  {M.}~\bibnamefont {Loewenhaupt}}, \bibinfo {author} {\bibfnamefont
  {C.}~\bibnamefont {Geibel}},\ and\ \bibinfo {author} {\bibfnamefont
  {F.}~\bibnamefont {Steglich}},\ }\href
  {https://doi.org/10.1103/PhysRevLett.92.136401} {\bibfield  {journal}
  {\bibinfo  {journal} {Phys. Rev. Lett.}\ }\textbf {\bibinfo {volume} {92}},\
  \bibinfo {pages} {136401} (\bibinfo {year} {2004})}\BibitemShut {NoStop}%
\bibitem [{\citenamefont {Arndt}\ \emph {et~al.}(2011)\citenamefont {Arndt},
  \citenamefont {Stockert}, \citenamefont {Schmalzl}, \citenamefont
  {Faulhaber}, \citenamefont {Jeevan}, \citenamefont {Geibel}, \citenamefont
  {Schmidt}, \citenamefont {Loewenhaupt},\ and\ \citenamefont
  {Steglich}}]{Arndt2011}%
  \BibitemOpen
  \bibfield  {author} {\bibinfo {author} {\bibfnamefont {J.}~\bibnamefont
  {Arndt}}, \bibinfo {author} {\bibfnamefont {O.}~\bibnamefont {Stockert}},
  \bibinfo {author} {\bibfnamefont {K.}~\bibnamefont {Schmalzl}}, \bibinfo
  {author} {\bibfnamefont {E.}~\bibnamefont {Faulhaber}}, \bibinfo {author}
  {\bibfnamefont {H.~S.}\ \bibnamefont {Jeevan}}, \bibinfo {author}
  {\bibfnamefont {C.}~\bibnamefont {Geibel}}, \bibinfo {author} {\bibfnamefont
  {W.}~\bibnamefont {Schmidt}}, \bibinfo {author} {\bibfnamefont
  {M.}~\bibnamefont {Loewenhaupt}},\ and\ \bibinfo {author} {\bibfnamefont
  {F.}~\bibnamefont {Steglich}},\ }\href
  {https://doi.org/10.1103/PhysRevLett.106.246401} {\bibfield  {journal}
  {\bibinfo  {journal} {Phys. Rev. Lett.}\ }\textbf {\bibinfo {volume} {106}},\
  \bibinfo {pages} {246401} (\bibinfo {year} {2011})}\BibitemShut {NoStop}%
\bibitem [{\citenamefont {Wu}\ \emph {et~al.}(2021{\natexlab{a}})\citenamefont
  {Wu}, \citenamefont {Fang}, \citenamefont {Su}, \citenamefont {Xie},
  \citenamefont {Li}, \citenamefont {Wu}, \citenamefont {Huang}, \citenamefont
  {Shen}, \citenamefont {Thiagarajan}, \citenamefont {Adell}, \citenamefont
  {Cao}, \citenamefont {Yuan}, \citenamefont {Steglich},\ and\ \citenamefont
  {Liu}}]{Wu2021CeCu2Si2}%
  \BibitemOpen
  \bibfield  {author} {\bibinfo {author} {\bibfnamefont {Z.}~\bibnamefont
  {Wu}}, \bibinfo {author} {\bibfnamefont {Y.}~\bibnamefont {Fang}}, \bibinfo
  {author} {\bibfnamefont {H.}~\bibnamefont {Su}}, \bibinfo {author}
  {\bibfnamefont {W.}~\bibnamefont {Xie}}, \bibinfo {author} {\bibfnamefont
  {P.}~\bibnamefont {Li}}, \bibinfo {author} {\bibfnamefont {Y.}~\bibnamefont
  {Wu}}, \bibinfo {author} {\bibfnamefont {Y.}~\bibnamefont {Huang}}, \bibinfo
  {author} {\bibfnamefont {D.}~\bibnamefont {Shen}}, \bibinfo {author}
  {\bibfnamefont {B.}~\bibnamefont {Thiagarajan}}, \bibinfo {author}
  {\bibfnamefont {J.}~\bibnamefont {Adell}}, \bibinfo {author} {\bibfnamefont
  {C.}~\bibnamefont {Cao}}, \bibinfo {author} {\bibfnamefont {H.-Q.}\
  \bibnamefont {Yuan}}, \bibinfo {author} {\bibfnamefont {F.}~\bibnamefont
  {Steglich}},\ and\ \bibinfo {author} {\bibfnamefont {Y.}~\bibnamefont
  {Liu}},\ }\href {https://doi.org/10.1103/PhysRevLett.127.067002} {\bibfield
  {journal} {\bibinfo  {journal} {Phys. Rev. Lett.}\ }\textbf {\bibinfo
  {volume} {127}},\ \bibinfo {pages} {067002} (\bibinfo {year}
  {2021}{\natexlab{a}})}\BibitemShut {NoStop}%
\bibitem [{\citenamefont {Si}\ \emph {et~al.}(2001)\citenamefont {Si},
  \citenamefont {Rabello}, \citenamefont {Ingersent},\ and\ \citenamefont
  {Smith}}]{Si2001Locally}%
  \BibitemOpen
  \bibfield  {author} {\bibinfo {author} {\bibfnamefont {Q.}~\bibnamefont
  {Si}}, \bibinfo {author} {\bibfnamefont {S.}~\bibnamefont {Rabello}},
  \bibinfo {author} {\bibfnamefont {K.}~\bibnamefont {Ingersent}},\ and\
  \bibinfo {author} {\bibfnamefont {J.~L.}\ \bibnamefont {Smith}},\ }\href
  {https://doi.org/10.1038/35101507} {\bibfield  {journal} {\bibinfo  {journal}
  {Nature}\ }\textbf {\bibinfo {volume} {413}},\ \bibinfo {pages} {804}
  (\bibinfo {year} {2001})}\BibitemShut {NoStop}%
\bibitem [{\citenamefont {Coleman}\ \emph {et~al.}(2001)\citenamefont
  {Coleman}, \citenamefont {P\'epin}, \citenamefont {Si},\ and\ \citenamefont
  {Ramazashvili}}]{Coleman2001How}%
  \BibitemOpen
  \bibfield  {author} {\bibinfo {author} {\bibfnamefont {P.}~\bibnamefont
  {Coleman}}, \bibinfo {author} {\bibfnamefont {C.}~\bibnamefont {P\'epin}},
  \bibinfo {author} {\bibfnamefont {Q.}~\bibnamefont {Si}},\ and\ \bibinfo
  {author} {\bibfnamefont {R.}~\bibnamefont {Ramazashvili}},\ }\href
  {https://doi.org/10.1088/0953-8984/13/35/202} {\bibfield  {journal} {\bibinfo
   {journal} {Journal of Physics: Condensed Matter}\ }\textbf {\bibinfo
  {volume} {13}},\ \bibinfo {pages} {R723} (\bibinfo {year}
  {2001})}\BibitemShut {NoStop}%
\bibitem [{\citenamefont {Custers}\ \emph {et~al.}(2003)\citenamefont
  {Custers}, \citenamefont {Gegenwart}, \citenamefont {Wilhelm}, \citenamefont
  {Neumaier}, \citenamefont {Tokiwa}, \citenamefont {Trovarelli}, \citenamefont
  {Geibel}, \citenamefont {Steglich}, \citenamefont {P\'epin},\ and\
  \citenamefont {Coleman}}]{Custers2003nature}%
  \BibitemOpen
  \bibfield  {author} {\bibinfo {author} {\bibfnamefont {J.}~\bibnamefont
  {Custers}}, \bibinfo {author} {\bibfnamefont {P.}~\bibnamefont {Gegenwart}},
  \bibinfo {author} {\bibfnamefont {H.}~\bibnamefont {Wilhelm}}, \bibinfo
  {author} {\bibfnamefont {K.}~\bibnamefont {Neumaier}}, \bibinfo {author}
  {\bibfnamefont {Y.}~\bibnamefont {Tokiwa}}, \bibinfo {author} {\bibfnamefont
  {O.}~\bibnamefont {Trovarelli}}, \bibinfo {author} {\bibfnamefont
  {C.}~\bibnamefont {Geibel}}, \bibinfo {author} {\bibfnamefont
  {F.}~\bibnamefont {Steglich}}, \bibinfo {author} {\bibfnamefont
  {C.}~\bibnamefont {P\'epin}},\ and\ \bibinfo {author} {\bibfnamefont
  {P.}~\bibnamefont {Coleman}},\ }\href {https://doi.org/10.1038/nature01774}
  {\bibfield  {journal} {\bibinfo  {journal} {Nature}\ }\textbf {\bibinfo
  {volume} {424}},\ \bibinfo {pages} {524} (\bibinfo {year}
  {2003})}\BibitemShut {NoStop}%
\bibitem [{\citenamefont {Shishido}\ \emph {et~al.}(2005)\citenamefont
  {Shishido}, \citenamefont {Settai}, \citenamefont {Harima},\ and\
  \citenamefont {\={O}nuki}}]{shishido2005JPSJ}%
  \BibitemOpen
  \bibfield  {author} {\bibinfo {author} {\bibfnamefont {H.}~\bibnamefont
  {Shishido}}, \bibinfo {author} {\bibfnamefont {R.}~\bibnamefont {Settai}},
  \bibinfo {author} {\bibfnamefont {H.}~\bibnamefont {Harima}},\ and\ \bibinfo
  {author} {\bibfnamefont {Y.}~\bibnamefont {\={O}nuki}},\ }\href
  {https://doi.org/10.1143/JPSJ.74.1103} {\bibfield  {journal} {\bibinfo
  {journal} {Journal of the Physical Society of Japan}\ }\textbf {\bibinfo
  {volume} {74}},\ \bibinfo {pages} {1103} (\bibinfo {year}
  {2005})}\BibitemShut {NoStop}%
\bibitem [{\citenamefont {Park}\ \emph {et~al.}(2006)\citenamefont {Park},
  \citenamefont {Ronning}, \citenamefont {Yuan}, \citenamefont {Salamon},
  \citenamefont {Movshovich}, \citenamefont {Sarrao},\ and\ \citenamefont
  {Thompson}}]{Park2006CeRhIn5}%
  \BibitemOpen
  \bibfield  {author} {\bibinfo {author} {\bibfnamefont {T.}~\bibnamefont
  {Park}}, \bibinfo {author} {\bibfnamefont {F.}~\bibnamefont {Ronning}},
  \bibinfo {author} {\bibfnamefont {H.-Q.}\ \bibnamefont {Yuan}}, \bibinfo
  {author} {\bibfnamefont {M.~B.}\ \bibnamefont {Salamon}}, \bibinfo {author}
  {\bibfnamefont {R.}~\bibnamefont {Movshovich}}, \bibinfo {author}
  {\bibfnamefont {J.~L.}\ \bibnamefont {Sarrao}},\ and\ \bibinfo {author}
  {\bibfnamefont {J.~D.}\ \bibnamefont {Thompson}},\ }\href
  {https://doi.org/10.1038/nature04571} {\bibfield  {journal} {\bibinfo
  {journal} {Nature}\ }\textbf {\bibinfo {volume} {440}},\ \bibinfo {pages}
  {65} (\bibinfo {year} {2006})}\BibitemShut {NoStop}%
\bibitem [{\citenamefont {Friedemann}\ \emph {et~al.}(2010)\citenamefont
  {Friedemann}, \citenamefont {Oeschler}, \citenamefont {Wirth}, \citenamefont
  {Krellner}, \citenamefont {Geibel}, \citenamefont {Steglich}, \citenamefont
  {Paschen}, \citenamefont {Kirschner},\ and\ \citenamefont
  {Si}}]{Friedeman2010PNAS}%
  \BibitemOpen
  \bibfield  {author} {\bibinfo {author} {\bibfnamefont {S.}~\bibnamefont
  {Friedemann}}, \bibinfo {author} {\bibfnamefont {N.}~\bibnamefont
  {Oeschler}}, \bibinfo {author} {\bibfnamefont {S.}~\bibnamefont {Wirth}},
  \bibinfo {author} {\bibfnamefont {C.}~\bibnamefont {Krellner}}, \bibinfo
  {author} {\bibfnamefont {C.}~\bibnamefont {Geibel}}, \bibinfo {author}
  {\bibfnamefont {F.}~\bibnamefont {Steglich}}, \bibinfo {author}
  {\bibfnamefont {S.}~\bibnamefont {Paschen}}, \bibinfo {author} {\bibfnamefont
  {S.}~\bibnamefont {Kirschner}},\ and\ \bibinfo {author} {\bibfnamefont
  {Q.}~\bibnamefont {Si}},\ }\href {https://doi.org/10.1073/pnas.1009202107}
  {\bibfield  {journal} {\bibinfo  {journal} {Proc. Natl. Acad. Sci.}\ }\textbf
  {\bibinfo {volume} {107}},\ \bibinfo {pages} {14547} (\bibinfo {year}
  {2010})}\BibitemShut {NoStop}%
\bibitem [{\citenamefont {Bork}\ \emph {et~al.}(2011)\citenamefont {Bork},
  \citenamefont {Zhang}, \citenamefont {Diekh\"oner}, \citenamefont {Borda},
  \citenamefont {Simon}, \citenamefont {Kroha}, \citenamefont {Wahl},\ and\
  \citenamefont {Kern}}]{Bork2011tunable}%
  \BibitemOpen
  \bibfield  {author} {\bibinfo {author} {\bibfnamefont {J.}~\bibnamefont
  {Bork}}, \bibinfo {author} {\bibfnamefont {Y.-h.}\ \bibnamefont {Zhang}},
  \bibinfo {author} {\bibfnamefont {L.}~\bibnamefont {Diekh\"oner}}, \bibinfo
  {author} {\bibfnamefont {L.}~\bibnamefont {Borda}}, \bibinfo {author}
  {\bibfnamefont {P.}~\bibnamefont {Simon}}, \bibinfo {author} {\bibfnamefont
  {J.}~\bibnamefont {Kroha}}, \bibinfo {author} {\bibfnamefont
  {P.}~\bibnamefont {Wahl}},\ and\ \bibinfo {author} {\bibfnamefont
  {K.}~\bibnamefont {Kern}},\ }\href {https://doi.org/10.1038/NPHYS2076}
  {\bibfield  {journal} {\bibinfo  {journal} {Nature Physics}\ }\textbf
  {\bibinfo {volume} {7}},\ \bibinfo {pages} {901} (\bibinfo {year}
  {2011})}\BibitemShut {NoStop}%
\bibitem [{\citenamefont {Pr\"{u}ser}\ \emph {et~al.}(2014)\citenamefont
  {Pr\"{u}ser}, \citenamefont {Dargel}, \citenamefont {Bouhassoune},
  \citenamefont {Ulbrich}, \citenamefont {Pruschke}, \citenamefont {Lounis},\
  and\ \citenamefont {Wenderoth}}]{Pruser2014interplay}%
  \BibitemOpen
  \bibfield  {author} {\bibinfo {author} {\bibfnamefont {H.}~\bibnamefont
  {Pr\"{u}ser}}, \bibinfo {author} {\bibfnamefont {P.~E.}\ \bibnamefont
  {Dargel}}, \bibinfo {author} {\bibfnamefont {M.}~\bibnamefont {Bouhassoune}},
  \bibinfo {author} {\bibfnamefont {R.~G.}\ \bibnamefont {Ulbrich}}, \bibinfo
  {author} {\bibfnamefont {T.}~\bibnamefont {Pruschke}}, \bibinfo {author}
  {\bibfnamefont {S.}~\bibnamefont {Lounis}},\ and\ \bibinfo {author}
  {\bibfnamefont {M.}~\bibnamefont {Wenderoth}},\ }\href
  {https://doi.org/10.1038/ncomms6417} {\bibfield  {journal} {\bibinfo
  {journal} {Nature Communications}\ }\textbf {\bibinfo {volume} {5}},\
  \bibinfo {pages} {5417} (\bibinfo {year} {2014})}\BibitemShut {NoStop}%
\bibitem [{\citenamefont {Poelchen}\ \emph {et~al.}(2020)\citenamefont
  {Poelchen}, \citenamefont {Schulz}, \citenamefont {Mende}, \citenamefont
  {G\"uttler}, \citenamefont {Generalov}, \citenamefont {Fedorov},
  \citenamefont {Caroca-Canales}, \citenamefont {Geibel}, \citenamefont
  {Kliemt}, \citenamefont {Krellner}, \citenamefont {Danzenb\"acher},
  \citenamefont {Yu.~Usachov}, \citenamefont {Dudin}, \citenamefont {Antonov},
  \citenamefont {Allen}, \citenamefont {Laubschat}, \citenamefont {Kummer},
  \citenamefont {Kucherenko},\ and\ \citenamefont
  {Vyalikh}}]{Poelchen2020unexpected}%
  \BibitemOpen
  \bibfield  {author} {\bibinfo {author} {\bibfnamefont {G.}~\bibnamefont
  {Poelchen}}, \bibinfo {author} {\bibfnamefont {S.}~\bibnamefont {Schulz}},
  \bibinfo {author} {\bibfnamefont {M.}~\bibnamefont {Mende}}, \bibinfo
  {author} {\bibfnamefont {M.}~\bibnamefont {G\"uttler}}, \bibinfo {author}
  {\bibfnamefont {A.}~\bibnamefont {Generalov}}, \bibinfo {author}
  {\bibfnamefont {A.~V.}\ \bibnamefont {Fedorov}}, \bibinfo {author}
  {\bibfnamefont {N.}~\bibnamefont {Caroca-Canales}}, \bibinfo {author}
  {\bibfnamefont {C.}~\bibnamefont {Geibel}}, \bibinfo {author} {\bibfnamefont
  {K.}~\bibnamefont {Kliemt}}, \bibinfo {author} {\bibfnamefont
  {C.}~\bibnamefont {Krellner}}, \bibinfo {author} {\bibfnamefont
  {S.}~\bibnamefont {Danzenb\"acher}}, \bibinfo {author} {\bibfnamefont
  {D.}~\bibnamefont {Yu.~Usachov}}, \bibinfo {author} {\bibfnamefont
  {P.}~\bibnamefont {Dudin}}, \bibinfo {author} {\bibfnamefont {V.~N.}\
  \bibnamefont {Antonov}}, \bibinfo {author} {\bibfnamefont {J.~W.}\
  \bibnamefont {Allen}}, \bibinfo {author} {\bibfnamefont {C.}~\bibnamefont
  {Laubschat}}, \bibinfo {author} {\bibfnamefont {K.}~\bibnamefont {Kummer}},
  \bibinfo {author} {\bibfnamefont {Y.}~\bibnamefont {Kucherenko}},\ and\
  \bibinfo {author} {\bibfnamefont {D.~V.}\ \bibnamefont {Vyalikh}},\ }\href
  {https://doi.org/10.1038/s41535-020-00273-7} {\bibfield  {journal} {\bibinfo
  {journal} {npj Quantum Materials}\ }\textbf {\bibinfo {volume} {5}},\
  \bibinfo {pages} {70} (\bibinfo {year} {2020})}\BibitemShut {NoStop}%
\bibitem [{\citenamefont {Pecharsky}\ \emph {et~al.}(1993)\citenamefont
  {Pecharsky}, \citenamefont {Hyun},\ and\ \citenamefont
  {K.~A.~Gschneidner}}]{Pecharsky1993PRB}%
  \BibitemOpen
  \bibfield  {author} {\bibinfo {author} {\bibfnamefont {V.~K.}\ \bibnamefont
  {Pecharsky}}, \bibinfo {author} {\bibfnamefont {O.-B.}\ \bibnamefont
  {Hyun}},\ and\ \bibinfo {author} {\bibfnamefont {J.}~\bibnamefont
  {K.~A.~Gschneidner}},\ }\href {https://doi.org/10.1103/physrevb.47.11839}
  {\bibfield  {journal} {\bibinfo  {journal} {Phys. Rev. B}\ }\textbf {\bibinfo
  {volume} {47}},\ \bibinfo {pages} {11839} (\bibinfo {year}
  {1993})}\BibitemShut {NoStop}%
\bibitem [{\citenamefont {Thamizhavel}\ \emph {et~al.}(2005)\citenamefont
  {Thamizhavel}, \citenamefont {Takeuchi}, \citenamefont {Matsuda},
  \citenamefont {Haga}, \citenamefont {Sugiyama}, \citenamefont {Settai},\ and\
  \citenamefont {\={O}nuki}}]{Thamizhavel2005JPSJ}%
  \BibitemOpen
  \bibfield  {author} {\bibinfo {author} {\bibfnamefont {A.}~\bibnamefont
  {Thamizhavel}}, \bibinfo {author} {\bibfnamefont {T.}~\bibnamefont
  {Takeuchi}}, \bibinfo {author} {\bibfnamefont {T.~D.}\ \bibnamefont
  {Matsuda}}, \bibinfo {author} {\bibfnamefont {Y.}~\bibnamefont {Haga}},
  \bibinfo {author} {\bibfnamefont {K.}~\bibnamefont {Sugiyama}}, \bibinfo
  {author} {\bibfnamefont {R.}~\bibnamefont {Settai}},\ and\ \bibinfo {author}
  {\bibfnamefont {Y.}~\bibnamefont {\={O}nuki}},\ }\href
  {https://doi.org/10.1143/jpsj.74.1858} {\bibfield  {journal} {\bibinfo
  {journal} {J. Phys. Soc. Jpn.}\ }\textbf {\bibinfo {volume} {74}},\ \bibinfo
  {pages} {1858} (\bibinfo {year} {2005})}\BibitemShut {NoStop}%
\bibitem [{\citenamefont {Smidman}\ \emph {et~al.}(2013)\citenamefont
  {Smidman}, \citenamefont {Adroja}, \citenamefont {Hillier}, \citenamefont
  {Chapon}, \citenamefont {Taylor}, \citenamefont {Anand}, \citenamefont
  {Singh}, \citenamefont {Lees}, \citenamefont {Goremychkin}, \citenamefont
  {Koza}, \citenamefont {Krishnamurthy}, \citenamefont {Paul},\ and\
  \citenamefont {Balakrishnan}}]{Michel2013neutronPRB}%
  \BibitemOpen
  \bibfield  {author} {\bibinfo {author} {\bibfnamefont {M.}~\bibnamefont
  {Smidman}}, \bibinfo {author} {\bibfnamefont {D.~T.}\ \bibnamefont {Adroja}},
  \bibinfo {author} {\bibfnamefont {A.~D.}\ \bibnamefont {Hillier}}, \bibinfo
  {author} {\bibfnamefont {L.~C.}\ \bibnamefont {Chapon}}, \bibinfo {author}
  {\bibfnamefont {J.~W.}\ \bibnamefont {Taylor}}, \bibinfo {author}
  {\bibfnamefont {V.~K.}\ \bibnamefont {Anand}}, \bibinfo {author}
  {\bibfnamefont {R.~P.}\ \bibnamefont {Singh}}, \bibinfo {author}
  {\bibfnamefont {M.~R.}\ \bibnamefont {Lees}}, \bibinfo {author}
  {\bibfnamefont {E.~A.}\ \bibnamefont {Goremychkin}}, \bibinfo {author}
  {\bibfnamefont {M.~M.}\ \bibnamefont {Koza}}, \bibinfo {author}
  {\bibfnamefont {V.~V.}\ \bibnamefont {Krishnamurthy}}, \bibinfo {author}
  {\bibfnamefont {D.~M.}\ \bibnamefont {Paul}},\ and\ \bibinfo {author}
  {\bibfnamefont {G.}~\bibnamefont {Balakrishnan}},\ }\href
  {https://doi.org/10.1103/PhysRevB.88.134416} {\bibfield  {journal} {\bibinfo
  {journal} {Phys. Rev. B}\ }\textbf {\bibinfo {volume} {88}},\ \bibinfo
  {pages} {134416} (\bibinfo {year} {2013})}\BibitemShut {NoStop}%
\bibitem [{\citenamefont {Ivanov}\ \emph {et~al.}(2021)\citenamefont {Ivanov},
  \citenamefont {Wan},\ and\ \citenamefont {Savrasov}}]{Xiangangwan2021cal}%
  \BibitemOpen
  \bibfield  {author} {\bibinfo {author} {\bibfnamefont {V.}~\bibnamefont
  {Ivanov}}, \bibinfo {author} {\bibfnamefont {X.}~\bibnamefont {Wan}},\ and\
  \bibinfo {author} {\bibfnamefont {S.~Y.}\ \bibnamefont {Savrasov}},\ }\href
  {https://doi.org/10.1103/PhysRevB.103.L041112} {\bibfield  {journal}
  {\bibinfo  {journal} {Phys. Rev. B}\ }\textbf {\bibinfo {volume} {103}},\
  \bibinfo {pages} {L041112} (\bibinfo {year} {2021})}\BibitemShut {NoStop}%
\bibitem [{\citenamefont {Settai}\ \emph {et~al.}(2007)\citenamefont {Settai},
  \citenamefont {Sugitani}, \citenamefont {Okuda}, \citenamefont {Thamizhavel},
  \citenamefont {Nakashima}, \citenamefont {\={O}nuki},\ and\ \citenamefont
  {Harima}}]{Settai2007JMMM}%
  \BibitemOpen
  \bibfield  {author} {\bibinfo {author} {\bibfnamefont {R.}~\bibnamefont
  {Settai}}, \bibinfo {author} {\bibfnamefont {I.}~\bibnamefont {Sugitani}},
  \bibinfo {author} {\bibfnamefont {Y.}~\bibnamefont {Okuda}}, \bibinfo
  {author} {\bibfnamefont {A.}~\bibnamefont {Thamizhavel}}, \bibinfo {author}
  {\bibfnamefont {M.}~\bibnamefont {Nakashima}}, \bibinfo {author}
  {\bibfnamefont {Y.}~\bibnamefont {\={O}nuki}},\ and\ \bibinfo {author}
  {\bibfnamefont {H.}~\bibnamefont {Harima}},\ }\href
  {https://doi.org/10.1016/j.jmmm.2006.10.717} {\bibfield  {journal} {\bibinfo
  {journal} {J. Magn. Magn. Mater.}\ }\textbf {\bibinfo {volume} {310}},\
  \bibinfo {pages} {844} (\bibinfo {year} {2007})}\BibitemShut {NoStop}%
\bibitem [{\citenamefont {Frigeri}\ \emph {et~al.}(2004)\citenamefont
  {Frigeri}, \citenamefont {Agterberg}, \citenamefont {Koga},\ and\
  \citenamefont {Sigrist}}]{Frigeri2004}%
  \BibitemOpen
  \bibfield  {author} {\bibinfo {author} {\bibfnamefont {P.~A.}\ \bibnamefont
  {Frigeri}}, \bibinfo {author} {\bibfnamefont {D.~F.}\ \bibnamefont
  {Agterberg}}, \bibinfo {author} {\bibfnamefont {A.}~\bibnamefont {Koga}},\
  and\ \bibinfo {author} {\bibfnamefont {M.}~\bibnamefont {Sigrist}},\ }\href
  {https://doi.org/10.1103/PhysRevLett.92.097001} {\bibfield  {journal}
  {\bibinfo  {journal} {Phys. Rev. Lett.}\ }\textbf {\bibinfo {volume} {92}},\
  \bibinfo {pages} {097001} (\bibinfo {year} {2004})}\BibitemShut {NoStop}%
\bibitem [{sup()}]{supplementary}%
  \BibitemOpen
  \href@noop {} {}\bibinfo {note} {See online supplementary material at xxx,
  which includes Refs.[34, 40-44], as well as ARPES measurement details,
  additional experimental data and analysis.}\BibitemShut {Stop}%
\bibitem [{\citenamefont {Yang}\ \emph {et~al.}(2021)\citenamefont {Yang},
  \citenamefont {Liu}, \citenamefont {Liu}, \citenamefont {Liu}, \citenamefont
  {Liu}, \citenamefont {Lu}, \citenamefont {Mei}, \citenamefont {Li},
  \citenamefont {Ye}, \citenamefont {Qiao},\ and\ \citenamefont
  {Shen}}]{Yang2021NST}%
  \BibitemOpen
  \bibfield  {author} {\bibinfo {author} {\bibfnamefont {Y.-C.}\ \bibnamefont
  {Yang}}, \bibinfo {author} {\bibfnamefont {Z.-T.}\ \bibnamefont {Liu}},
  \bibinfo {author} {\bibfnamefont {J.-S.}\ \bibnamefont {Liu}}, \bibinfo
  {author} {\bibfnamefont {Z.-H.}\ \bibnamefont {Liu}}, \bibinfo {author}
  {\bibfnamefont {W.-L.}\ \bibnamefont {Liu}}, \bibinfo {author} {\bibfnamefont
  {X.-L.}\ \bibnamefont {Lu}}, \bibinfo {author} {\bibfnamefont {H.-P.}\
  \bibnamefont {Mei}}, \bibinfo {author} {\bibfnamefont {A.}~\bibnamefont
  {Li}}, \bibinfo {author} {\bibfnamefont {M.}~\bibnamefont {Ye}}, \bibinfo
  {author} {\bibfnamefont {S.}~\bibnamefont {Qiao}},\ and\ \bibinfo {author}
  {\bibfnamefont {D.-W.}\ \bibnamefont {Shen}},\ }\href
  {https://doi.org/10.1007/s41365-021-00858-2} {\bibfield  {journal} {\bibinfo
  {journal} {Nuc. Sci. Tech.}\ }\textbf {\bibinfo {volume} {32}},\ \bibinfo
  {pages} {31} (\bibinfo {year} {2021})}\BibitemShut {NoStop}%
\bibitem [{\citenamefont {Kresse}\ and\ \citenamefont
  {Hafner}(1993)}]{PhysRevB.47.558}%
  \BibitemOpen
  \bibfield  {author} {\bibinfo {author} {\bibfnamefont {G.}~\bibnamefont
  {Kresse}}\ and\ \bibinfo {author} {\bibfnamefont {J.}~\bibnamefont
  {Hafner}},\ }\href {https://doi.org/10.1103/PhysRevB.47.558} {\bibfield
  {journal} {\bibinfo  {journal} {Phys. Rev. B}\ }\textbf {\bibinfo {volume}
  {47}},\ \bibinfo {pages} {558} (\bibinfo {year} {1993})}\BibitemShut
  {NoStop}%
\bibitem [{\citenamefont {Kresse}\ and\ \citenamefont
  {Joubert}(1999)}]{PhysRevB.59.1758}%
  \BibitemOpen
  \bibfield  {author} {\bibinfo {author} {\bibfnamefont {G.}~\bibnamefont
  {Kresse}}\ and\ \bibinfo {author} {\bibfnamefont {D.}~\bibnamefont
  {Joubert}},\ }\href {https://doi.org/10.1103/PhysRevB.59.1758} {\bibfield
  {journal} {\bibinfo  {journal} {Phys. Rev. B}\ }\textbf {\bibinfo {volume}
  {59}},\ \bibinfo {pages} {1758} (\bibinfo {year} {1999})}\BibitemShut
  {NoStop}%
\bibitem [{\citenamefont {Perdew}\ \emph {et~al.}(1996)\citenamefont {Perdew},
  \citenamefont {Burke},\ and\ \citenamefont
  {Ernzerhof}}]{PhysRevLett.77.3865}%
  \BibitemOpen
  \bibfield  {author} {\bibinfo {author} {\bibfnamefont {J.~P.}\ \bibnamefont
  {Perdew}}, \bibinfo {author} {\bibfnamefont {K.}~\bibnamefont {Burke}},\ and\
  \bibinfo {author} {\bibfnamefont {M.}~\bibnamefont {Ernzerhof}},\ }\href
  {https://doi.org/10.1103/PhysRevLett.77.3865} {\bibfield  {journal} {\bibinfo
   {journal} {Phys. Rev. Lett.}\ }\textbf {\bibinfo {volume} {77}},\ \bibinfo
  {pages} {3865} (\bibinfo {year} {1996})}\BibitemShut {NoStop}%
\bibitem [{\citenamefont {Mostofi}\ \emph {et~al.}(2008)\citenamefont
  {Mostofi}, \citenamefont {Yates}, \citenamefont {Lee}, \citenamefont {Souza},
  \citenamefont {Vanderbilt},\ and\ \citenamefont {Marzari}}]{MOSTOFI2008685}%
  \BibitemOpen
  \bibfield  {author} {\bibinfo {author} {\bibfnamefont {A.~A.}\ \bibnamefont
  {Mostofi}}, \bibinfo {author} {\bibfnamefont {J.~R.}\ \bibnamefont {Yates}},
  \bibinfo {author} {\bibfnamefont {Y.-S.}\ \bibnamefont {Lee}}, \bibinfo
  {author} {\bibfnamefont {I.}~\bibnamefont {Souza}}, \bibinfo {author}
  {\bibfnamefont {D.}~\bibnamefont {Vanderbilt}},\ and\ \bibinfo {author}
  {\bibfnamefont {N.}~\bibnamefont {Marzari}},\ }\href
  {https://doi.org/https://doi.org/10.1016/j.cpc.2007.11.016} {\bibfield
  {journal} {\bibinfo  {journal} {Computer Physics Communications}\ }\textbf
  {\bibinfo {volume} {178}},\ \bibinfo {pages} {685 } (\bibinfo {year}
  {2008})}\BibitemShut {NoStop}%
\bibitem [{\citenamefont {Thamizhavel}\ \emph {et~al.}(2006)\citenamefont
  {Thamizhavel}, \citenamefont {Shishido}, \citenamefont {Okuda}, \citenamefont
  {Harima}, \citenamefont {Matsuda}, \citenamefont {Haga}, \citenamefont
  {Settai},\ and\ \citenamefont {\={O}nuki}}]{Thamizhavel2006JPSJ}%
  \BibitemOpen
  \bibfield  {author} {\bibinfo {author} {\bibfnamefont {A.}~\bibnamefont
  {Thamizhavel}}, \bibinfo {author} {\bibfnamefont {H.}~\bibnamefont
  {Shishido}}, \bibinfo {author} {\bibfnamefont {Y.}~\bibnamefont {Okuda}},
  \bibinfo {author} {\bibfnamefont {H.}~\bibnamefont {Harima}}, \bibinfo
  {author} {\bibfnamefont {T.~D.}\ \bibnamefont {Matsuda}}, \bibinfo {author}
  {\bibfnamefont {Y.}~\bibnamefont {Haga}}, \bibinfo {author} {\bibfnamefont
  {R.}~\bibnamefont {Settai}},\ and\ \bibinfo {author} {\bibfnamefont
  {Y.}~\bibnamefont {\={O}nuki}},\ }\href
  {https://doi.org/10.1143/jpsj.75.044711} {\bibfield  {journal} {\bibinfo
  {journal} {J. Phys. Soc. Jpn.}\ }\textbf {\bibinfo {volume} {75}},\ \bibinfo
  {pages} {044711} (\bibinfo {year} {2006})}\BibitemShut {NoStop}%
\bibitem [{\citenamefont {Allen}(2005)}]{jw2005kondo}%
  \BibitemOpen
  \bibfield  {author} {\bibinfo {author} {\bibfnamefont {J.~W.}\ \bibnamefont
  {Allen}},\ }\href {https://doi.org/10.1143/JPSJ.74.34} {\bibfield  {journal}
  {\bibinfo  {journal} {Journal of the Physical Society of Japan}\ }\textbf
  {\bibinfo {volume} {74}},\ \bibinfo {pages} {34} (\bibinfo {year}
  {2005})}\BibitemShut {NoStop}%
\bibitem [{\citenamefont {Sekiyama}\ \emph {et~al.}(2000)\citenamefont
  {Sekiyama}, \citenamefont {Iwasaki}, \citenamefont {Matsuda}, \citenamefont
  {Saitoh}, \citenamefont {\={O}nuki},\ and\ \citenamefont
  {Suga}}]{sekiyama2000probing}%
  \BibitemOpen
  \bibfield  {author} {\bibinfo {author} {\bibfnamefont {A.}~\bibnamefont
  {Sekiyama}}, \bibinfo {author} {\bibfnamefont {T.}~\bibnamefont {Iwasaki}},
  \bibinfo {author} {\bibfnamefont {K.}~\bibnamefont {Matsuda}}, \bibinfo
  {author} {\bibfnamefont {Y.}~\bibnamefont {Saitoh}}, \bibinfo {author}
  {\bibfnamefont {Y.}~\bibnamefont {\={O}nuki}},\ and\ \bibinfo {author}
  {\bibfnamefont {S.}~\bibnamefont {Suga}},\ }\href
  {https://doi.org/10.1038/35000140} {\bibfield  {journal} {\bibinfo  {journal}
  {Nature}\ }\textbf {\bibinfo {volume} {403}},\ \bibinfo {pages} {396}
  (\bibinfo {year} {2000})}\BibitemShut {NoStop}%
\bibitem [{\citenamefont {Im}\ \emph {et~al.}(2008)\citenamefont {Im},
  \citenamefont {Ito}, \citenamefont {Kim}, \citenamefont {Kimura},
  \citenamefont {Lee}, \citenamefont {Hong}, \citenamefont {Kwon},
  \citenamefont {Yasui},\ and\ \citenamefont {Yamagami}}]{Im2008}%
  \BibitemOpen
  \bibfield  {author} {\bibinfo {author} {\bibfnamefont {H.~J.}\ \bibnamefont
  {Im}}, \bibinfo {author} {\bibfnamefont {T.}~\bibnamefont {Ito}}, \bibinfo
  {author} {\bibfnamefont {H.-D.}\ \bibnamefont {Kim}}, \bibinfo {author}
  {\bibfnamefont {S.}~\bibnamefont {Kimura}}, \bibinfo {author} {\bibfnamefont
  {K.~E.}\ \bibnamefont {Lee}}, \bibinfo {author} {\bibfnamefont {J.~B.}\
  \bibnamefont {Hong}}, \bibinfo {author} {\bibfnamefont {Y.~S.}\ \bibnamefont
  {Kwon}}, \bibinfo {author} {\bibfnamefont {A.}~\bibnamefont {Yasui}},\ and\
  \bibinfo {author} {\bibfnamefont {H.}~\bibnamefont {Yamagami}},\ }\href
  {https://doi.org/10.1103/PhysRevLett.100.176402} {\bibfield  {journal}
  {\bibinfo  {journal} {Phys. Rev. Lett.}\ }\textbf {\bibinfo {volume} {100}},\
  \bibinfo {pages} {176402} (\bibinfo {year} {2008})}\BibitemShut {NoStop}%
\bibitem [{\citenamefont {Patil}\ \emph {et~al.}(2016)\citenamefont {Patil},
  \citenamefont {Generalov}, \citenamefont {G\"uttler}, \citenamefont
  {Kushwaha}, \citenamefont {Chikina}, \citenamefont {Kummer}, \citenamefont
  {R\"odel}, \citenamefont {Santander-Syro}, \citenamefont {Caroca-Canales},
  \citenamefont {Geibel}, \citenamefont {Danzenb\"acher}, \citenamefont
  {Kucherenko}, \citenamefont {Laubschat}, \citenamefont {Allen},\ and\
  \citenamefont {Vyalikh}}]{patil2016arpes}%
  \BibitemOpen
  \bibfield  {author} {\bibinfo {author} {\bibfnamefont {S.}~\bibnamefont
  {Patil}}, \bibinfo {author} {\bibfnamefont {A.}~\bibnamefont {Generalov}},
  \bibinfo {author} {\bibfnamefont {M.}~\bibnamefont {G\"uttler}}, \bibinfo
  {author} {\bibfnamefont {P.}~\bibnamefont {Kushwaha}}, \bibinfo {author}
  {\bibfnamefont {A.}~\bibnamefont {Chikina}}, \bibinfo {author} {\bibfnamefont
  {K.}~\bibnamefont {Kummer}}, \bibinfo {author} {\bibfnamefont {T.~C.}\
  \bibnamefont {R\"odel}}, \bibinfo {author} {\bibfnamefont {A.~F.}\
  \bibnamefont {Santander-Syro}}, \bibinfo {author} {\bibfnamefont
  {N.}~\bibnamefont {Caroca-Canales}}, \bibinfo {author} {\bibfnamefont
  {C.}~\bibnamefont {Geibel}}, \bibinfo {author} {\bibfnamefont
  {S.}~\bibnamefont {Danzenb\"acher}}, \bibinfo {author} {\bibfnamefont
  {Y.}~\bibnamefont {Kucherenko}}, \bibinfo {author} {\bibfnamefont
  {C.}~\bibnamefont {Laubschat}}, \bibinfo {author} {\bibfnamefont {J.~W.}\
  \bibnamefont {Allen}},\ and\ \bibinfo {author} {\bibfnamefont {D.~V.}\
  \bibnamefont {Vyalikh}},\ }\href {https://doi.org/10.1038/ncomms11029}
  {\bibfield  {journal} {\bibinfo  {journal} {Nature Communications}\ }\textbf
  {\bibinfo {volume} {7}},\ \bibinfo {pages} {1} (\bibinfo {year}
  {2016})}\BibitemShut {NoStop}%
\bibitem [{\citenamefont {Chen}\ \emph {et~al.}(2017)\citenamefont {Chen},
  \citenamefont {Xu}, \citenamefont {Niu}, \citenamefont {Jiang}, \citenamefont
  {Peng}, \citenamefont {Xu}, \citenamefont {Wen}, \citenamefont {Ding},
  \citenamefont {Huang}, \citenamefont {Shu}, \citenamefont {Zhang},
  \citenamefont {Lee}, \citenamefont {Strocov}, \citenamefont {Shi},
  \citenamefont {Bisti}, \citenamefont {Schmitt}, \citenamefont {Huang},
  \citenamefont {Dudin}, \citenamefont {Lai}, \citenamefont {Kirchner},
  \citenamefont {Yuan},\ and\ \citenamefont {Feng}}]{Chen2017}%
  \BibitemOpen
  \bibfield  {author} {\bibinfo {author} {\bibfnamefont {Q.~Y.}\ \bibnamefont
  {Chen}}, \bibinfo {author} {\bibfnamefont {D.~F.}\ \bibnamefont {Xu}},
  \bibinfo {author} {\bibfnamefont {X.~H.}\ \bibnamefont {Niu}}, \bibinfo
  {author} {\bibfnamefont {J.}~\bibnamefont {Jiang}}, \bibinfo {author}
  {\bibfnamefont {R.}~\bibnamefont {Peng}}, \bibinfo {author} {\bibfnamefont
  {H.~C.}\ \bibnamefont {Xu}}, \bibinfo {author} {\bibfnamefont {C.~H.~P.}\
  \bibnamefont {Wen}}, \bibinfo {author} {\bibfnamefont {Z.~F.}\ \bibnamefont
  {Ding}}, \bibinfo {author} {\bibfnamefont {K.}~\bibnamefont {Huang}},
  \bibinfo {author} {\bibfnamefont {L.}~\bibnamefont {Shu}}, \bibinfo {author}
  {\bibfnamefont {Y.~J.}\ \bibnamefont {Zhang}}, \bibinfo {author}
  {\bibfnamefont {H.}~\bibnamefont {Lee}}, \bibinfo {author} {\bibfnamefont
  {V.~N.}\ \bibnamefont {Strocov}}, \bibinfo {author} {\bibfnamefont
  {M.}~\bibnamefont {Shi}}, \bibinfo {author} {\bibfnamefont {F.}~\bibnamefont
  {Bisti}}, \bibinfo {author} {\bibfnamefont {T.}~\bibnamefont {Schmitt}},
  \bibinfo {author} {\bibfnamefont {Y.~B.}\ \bibnamefont {Huang}}, \bibinfo
  {author} {\bibfnamefont {P.}~\bibnamefont {Dudin}}, \bibinfo {author}
  {\bibfnamefont {X.~C.}\ \bibnamefont {Lai}}, \bibinfo {author} {\bibfnamefont
  {S.}~\bibnamefont {Kirchner}}, \bibinfo {author} {\bibfnamefont {H.~Q.}\
  \bibnamefont {Yuan}},\ and\ \bibinfo {author} {\bibfnamefont {D.~L.}\
  \bibnamefont {Feng}},\ }\href {https://doi.org/10.1103/PhysRevB.96.045107}
  {\bibfield  {journal} {\bibinfo  {journal} {Phys. Rev. B}\ }\textbf {\bibinfo
  {volume} {96}},\ \bibinfo {pages} {045107} (\bibinfo {year}
  {2017})}\BibitemShut {NoStop}%
\bibitem [{\citenamefont {Jang}\ \emph {et~al.}(2020)\citenamefont {Jang},
  \citenamefont {Denlinger}, \citenamefont {Allen}, \citenamefont {Zapf},
  \citenamefont {Maple}, \citenamefont {Kim}, \citenamefont {Jang},\ and\
  \citenamefont {Shim}}]{Jang2020}%
  \BibitemOpen
  \bibfield  {author} {\bibinfo {author} {\bibfnamefont {S.~Y.}\ \bibnamefont
  {Jang}}, \bibinfo {author} {\bibfnamefont {J.~D.}\ \bibnamefont {Denlinger}},
  \bibinfo {author} {\bibfnamefont {J.~W.}\ \bibnamefont {Allen}}, \bibinfo
  {author} {\bibfnamefont {V.~S.}\ \bibnamefont {Zapf}}, \bibinfo {author}
  {\bibfnamefont {M.~B.}\ \bibnamefont {Maple}}, \bibinfo {author}
  {\bibfnamefont {J.~N.}\ \bibnamefont {Kim}}, \bibinfo {author} {\bibfnamefont
  {B.~G.}\ \bibnamefont {Jang}},\ and\ \bibinfo {author} {\bibfnamefont
  {J.~H.}\ \bibnamefont {Shim}},\ }\href
  {https://doi.org/10.1073/pnas.2001778117} {\bibfield  {journal} {\bibinfo
  {journal} {Proc. Natl. Acad. Sci.}\ }\textbf {\bibinfo {volume} {117}},\
  \bibinfo {pages} {23467} (\bibinfo {year} {2020})}\BibitemShut {NoStop}%
\bibitem [{\citenamefont {Wu}\ \emph {et~al.}(2021{\natexlab{b}})\citenamefont
  {Wu}, \citenamefont {Zhang}, \citenamefont {Du}, \citenamefont {Shen},
  \citenamefont {Zheng}, \citenamefont {Fang}, \citenamefont {Smidman},
  \citenamefont {Cao}, \citenamefont {Steglich}, \citenamefont {Yuan},
  \citenamefont {Denlinger},\ and\ \citenamefont {Liu}}]{WuCeRh6Ge4PRL2021}%
  \BibitemOpen
  \bibfield  {author} {\bibinfo {author} {\bibfnamefont {Y.}~\bibnamefont
  {Wu}}, \bibinfo {author} {\bibfnamefont {Y.}~\bibnamefont {Zhang}}, \bibinfo
  {author} {\bibfnamefont {F.}~\bibnamefont {Du}}, \bibinfo {author}
  {\bibfnamefont {B.}~\bibnamefont {Shen}}, \bibinfo {author} {\bibfnamefont
  {H.}~\bibnamefont {Zheng}}, \bibinfo {author} {\bibfnamefont
  {Y.}~\bibnamefont {Fang}}, \bibinfo {author} {\bibfnamefont {M.}~\bibnamefont
  {Smidman}}, \bibinfo {author} {\bibfnamefont {C.}~\bibnamefont {Cao}},
  \bibinfo {author} {\bibfnamefont {F.}~\bibnamefont {Steglich}}, \bibinfo
  {author} {\bibfnamefont {H.}~\bibnamefont {Yuan}}, \bibinfo {author}
  {\bibfnamefont {J.~D.}\ \bibnamefont {Denlinger}},\ and\ \bibinfo {author}
  {\bibfnamefont {Y.}~\bibnamefont {Liu}},\ }\href
  {https://doi.org/10.1103/PhysRevLett.126.216406} {\bibfield  {journal}
  {\bibinfo  {journal} {Phys. Rev. Lett.}\ }\textbf {\bibinfo {volume} {126}},\
  \bibinfo {pages} {216406} (\bibinfo {year} {2021}{\natexlab{b}})}\BibitemShut
  {NoStop}%
\bibitem [{\citenamefont {Ehm}\ \emph {et~al.}(2007)\citenamefont {Ehm},
  \citenamefont {H\"ufner}, \citenamefont {Reinert}, \citenamefont {Kroha},
  \citenamefont {W\"olfle}, \citenamefont {Stockert}, \citenamefont {Geibel},\
  and\ \citenamefont {L\"ohneysen}}]{ehm2007high}%
  \BibitemOpen
  \bibfield  {author} {\bibinfo {author} {\bibfnamefont {D.}~\bibnamefont
  {Ehm}}, \bibinfo {author} {\bibfnamefont {S.}~\bibnamefont {H\"ufner}},
  \bibinfo {author} {\bibfnamefont {F.}~\bibnamefont {Reinert}}, \bibinfo
  {author} {\bibfnamefont {J.}~\bibnamefont {Kroha}}, \bibinfo {author}
  {\bibfnamefont {P.}~\bibnamefont {W\"olfle}}, \bibinfo {author}
  {\bibfnamefont {O.}~\bibnamefont {Stockert}}, \bibinfo {author}
  {\bibfnamefont {C.}~\bibnamefont {Geibel}},\ and\ \bibinfo {author}
  {\bibfnamefont {H.~v.}\ \bibnamefont {L\"ohneysen}},\ }\href
  {https://doi.org/10.1103/PhysRevB.76.045117} {\bibfield  {journal} {\bibinfo
  {journal} {Phys. Rev. B}\ }\textbf {\bibinfo {volume} {76}},\ \bibinfo
  {pages} {045117} (\bibinfo {year} {2007})}\BibitemShut {NoStop}%
\bibitem [{\citenamefont {Reinert}\ \emph {et~al.}(2001)\citenamefont
  {Reinert}, \citenamefont {Ehm}, \citenamefont {Schmidt}, \citenamefont
  {Nicolay}, \citenamefont {H\"ufner}, \citenamefont {Kroha}, \citenamefont
  {Trovarelli},\ and\ \citenamefont {Geibel}}]{Reinert2001}%
  \BibitemOpen
  \bibfield  {author} {\bibinfo {author} {\bibfnamefont {F.}~\bibnamefont
  {Reinert}}, \bibinfo {author} {\bibfnamefont {D.}~\bibnamefont {Ehm}},
  \bibinfo {author} {\bibfnamefont {S.}~\bibnamefont {Schmidt}}, \bibinfo
  {author} {\bibfnamefont {G.}~\bibnamefont {Nicolay}}, \bibinfo {author}
  {\bibfnamefont {S.}~\bibnamefont {H\"ufner}}, \bibinfo {author}
  {\bibfnamefont {J.}~\bibnamefont {Kroha}}, \bibinfo {author} {\bibfnamefont
  {O.}~\bibnamefont {Trovarelli}},\ and\ \bibinfo {author} {\bibfnamefont
  {C.}~\bibnamefont {Geibel}},\ }\href
  {https://doi.org/10.1103/PhysRevLett.87.106401} {\bibfield  {journal}
  {\bibinfo  {journal} {Phys. Rev. Lett.}\ }\textbf {\bibinfo {volume} {87}},\
  \bibinfo {pages} {106401} (\bibinfo {year} {2001})}\BibitemShut {NoStop}%
\bibitem [{\citenamefont {Kroha}\ \emph {et~al.}(2003)\citenamefont {Kroha},
  \citenamefont {Kirchner}, \citenamefont {Sellier}, \citenamefont {Wölfle},
  \citenamefont {Ehm}, \citenamefont {Reinert}, \citenamefont {Hüfner},\ and\
  \citenamefont {Geibel}}]{Kroha2003}%
  \BibitemOpen
  \bibfield  {author} {\bibinfo {author} {\bibfnamefont {J.}~\bibnamefont
  {Kroha}}, \bibinfo {author} {\bibfnamefont {S.}~\bibnamefont {Kirchner}},
  \bibinfo {author} {\bibfnamefont {G.}~\bibnamefont {Sellier}}, \bibinfo
  {author} {\bibfnamefont {P.}~\bibnamefont {Wölfle}}, \bibinfo {author}
  {\bibfnamefont {D.}~\bibnamefont {Ehm}}, \bibinfo {author} {\bibfnamefont
  {F.}~\bibnamefont {Reinert}}, \bibinfo {author} {\bibfnamefont
  {S.}~\bibnamefont {Hüfner}},\ and\ \bibinfo {author} {\bibfnamefont
  {C.}~\bibnamefont {Geibel}},\ }\href
  {https://doi.org/https://doi.org/10.1016/S1386-9477(02)00977-3} {\bibfield
  {journal} {\bibinfo  {journal} {Physica E: Low-dimensional Systems and
  Nanostructures}\ }\textbf {\bibinfo {volume} {18}},\ \bibinfo {pages} {69}
  (\bibinfo {year} {2003})},\ \bibinfo {note} {23rd International Conference on
  Low Temperature Physics (LT23)}\BibitemShut {NoStop}%
\bibitem [{\citenamefont {Luo}\ \emph {et~al.}(2020)\citenamefont {Luo},
  \citenamefont {Zhang}, \citenamefont {Wu}, \citenamefont {Wu}, \citenamefont
  {Song}, \citenamefont {Xia}, \citenamefont {Guo}, \citenamefont {Rusz},
  \citenamefont {Oppeneer}, \citenamefont {Durakiewicz}, \citenamefont {Zhao},
  \citenamefont {Liu}, \citenamefont {Zhu}, \citenamefont {Yuan}, \citenamefont
  {Tang}, \citenamefont {He}, \citenamefont {Tan}, \citenamefont {Huang},
  \citenamefont {Sun}, \citenamefont {Liu}, \citenamefont {Liu}, \citenamefont
  {Duan},\ and\ \citenamefont {Meng}}]{Luo2020PRB}%
  \BibitemOpen
  \bibfield  {author} {\bibinfo {author} {\bibfnamefont {Y.}~\bibnamefont
  {Luo}}, \bibinfo {author} {\bibfnamefont {C.}~\bibnamefont {Zhang}}, \bibinfo
  {author} {\bibfnamefont {Q.-Y.}\ \bibnamefont {Wu}}, \bibinfo {author}
  {\bibfnamefont {F.-Y.}\ \bibnamefont {Wu}}, \bibinfo {author} {\bibfnamefont
  {J.-J.}\ \bibnamefont {Song}}, \bibinfo {author} {\bibfnamefont
  {W.}~\bibnamefont {Xia}}, \bibinfo {author} {\bibfnamefont {Y.}~\bibnamefont
  {Guo}}, \bibinfo {author} {\bibfnamefont {J.}~\bibnamefont {Rusz}}, \bibinfo
  {author} {\bibfnamefont {P.~M.}\ \bibnamefont {Oppeneer}}, \bibinfo {author}
  {\bibfnamefont {T.}~\bibnamefont {Durakiewicz}}, \bibinfo {author}
  {\bibfnamefont {Y.-Z.}\ \bibnamefont {Zhao}}, \bibinfo {author}
  {\bibfnamefont {H.}~\bibnamefont {Liu}}, \bibinfo {author} {\bibfnamefont
  {S.-X.}\ \bibnamefont {Zhu}}, \bibinfo {author} {\bibfnamefont {Y.-H.}\
  \bibnamefont {Yuan}}, \bibinfo {author} {\bibfnamefont {X.-F.}\ \bibnamefont
  {Tang}}, \bibinfo {author} {\bibfnamefont {J.}~\bibnamefont {He}}, \bibinfo
  {author} {\bibfnamefont {S.-Y.}\ \bibnamefont {Tan}}, \bibinfo {author}
  {\bibfnamefont {Y.-B.}\ \bibnamefont {Huang}}, \bibinfo {author}
  {\bibfnamefont {Z.}~\bibnamefont {Sun}}, \bibinfo {author} {\bibfnamefont
  {Y.}~\bibnamefont {Liu}}, \bibinfo {author} {\bibfnamefont {H.~Y.}\
  \bibnamefont {Liu}}, \bibinfo {author} {\bibfnamefont {Y.-X.}\ \bibnamefont
  {Duan}},\ and\ \bibinfo {author} {\bibfnamefont {J.-Q.}\ \bibnamefont
  {Meng}},\ }\href {https://doi.org/10.1103/PhysRevB.101.115129} {\bibfield
  {journal} {\bibinfo  {journal} {Phys. Rev. B}\ }\textbf {\bibinfo {volume}
  {101}},\ \bibinfo {pages} {115129} (\bibinfo {year} {2020})}\BibitemShut
  {NoStop}%
\bibitem [{\citenamefont {apRoberts Warren}\ \emph {et~al.}(2011)\citenamefont
  {apRoberts Warren}, \citenamefont {Dioguardi}, \citenamefont {Shockley},
  \citenamefont {Lin}, \citenamefont {Crocker}, \citenamefont {Klavins},
  \citenamefont {Pines}, \citenamefont {Yang},\ and\ \citenamefont
  {Curro}}]{apRoberts2011PRB}%
  \BibitemOpen
  \bibfield  {author} {\bibinfo {author} {\bibfnamefont {N.}~\bibnamefont
  {apRoberts Warren}}, \bibinfo {author} {\bibfnamefont {A.~P.}\ \bibnamefont
  {Dioguardi}}, \bibinfo {author} {\bibfnamefont {A.~C.}\ \bibnamefont
  {Shockley}}, \bibinfo {author} {\bibfnamefont {C.~H.}\ \bibnamefont {Lin}},
  \bibinfo {author} {\bibfnamefont {J.}~\bibnamefont {Crocker}}, \bibinfo
  {author} {\bibfnamefont {P.}~\bibnamefont {Klavins}}, \bibinfo {author}
  {\bibfnamefont {D.}~\bibnamefont {Pines}}, \bibinfo {author} {\bibfnamefont
  {Y.-F.}\ \bibnamefont {Yang}},\ and\ \bibinfo {author} {\bibfnamefont
  {N.~J.}\ \bibnamefont {Curro}},\ }\href
  {https://doi.org/10.1103/PhysRevB.83.060408} {\bibfield  {journal} {\bibinfo
  {journal} {Phys. Rev. B}\ }\textbf {\bibinfo {volume} {83}},\ \bibinfo
  {pages} {060408(R)} (\bibinfo {year} {2011})}\BibitemShut {NoStop}%
\bibitem [{\citenamefont {Chen}\ \emph {et~al.}(2019)\citenamefont {Chen},
  \citenamefont {Luo}, \citenamefont {Xie}, \citenamefont {Li}, \citenamefont
  {Ji}, \citenamefont {Zhou}, \citenamefont {Huang}, \citenamefont {Zhang},
  \citenamefont {Feng}, \citenamefont {Zhang}, \citenamefont {Huang},
  \citenamefont {Hao}, \citenamefont {Liu}, \citenamefont {Zhu}, \citenamefont
  {Liu}, \citenamefont {Zhang}, \citenamefont {Lai}, \citenamefont {Si},\ and\
  \citenamefont {Tan}}]{Chen2019Orbital}%
  \BibitemOpen
  \bibfield  {author} {\bibinfo {author} {\bibfnamefont {Q.~Y.}\ \bibnamefont
  {Chen}}, \bibinfo {author} {\bibfnamefont {X.~B.}\ \bibnamefont {Luo}},
  \bibinfo {author} {\bibfnamefont {D.~H.}\ \bibnamefont {Xie}}, \bibinfo
  {author} {\bibfnamefont {M.~L.}\ \bibnamefont {Li}}, \bibinfo {author}
  {\bibfnamefont {X.~Y.}\ \bibnamefont {Ji}}, \bibinfo {author} {\bibfnamefont
  {R.}~\bibnamefont {Zhou}}, \bibinfo {author} {\bibfnamefont {Y.~B.}\
  \bibnamefont {Huang}}, \bibinfo {author} {\bibfnamefont {W.}~\bibnamefont
  {Zhang}}, \bibinfo {author} {\bibfnamefont {W.}~\bibnamefont {Feng}},
  \bibinfo {author} {\bibfnamefont {Y.}~\bibnamefont {Zhang}}, \bibinfo
  {author} {\bibfnamefont {L.}~\bibnamefont {Huang}}, \bibinfo {author}
  {\bibfnamefont {Q.~Q.}\ \bibnamefont {Hao}}, \bibinfo {author} {\bibfnamefont
  {Q.}~\bibnamefont {Liu}}, \bibinfo {author} {\bibfnamefont {X.~G.}\
  \bibnamefont {Zhu}}, \bibinfo {author} {\bibfnamefont {Y.}~\bibnamefont
  {Liu}}, \bibinfo {author} {\bibfnamefont {P.}~\bibnamefont {Zhang}}, \bibinfo
  {author} {\bibfnamefont {X.~C.}\ \bibnamefont {Lai}}, \bibinfo {author}
  {\bibfnamefont {Q.}~\bibnamefont {Si}},\ and\ \bibinfo {author}
  {\bibfnamefont {S.~Y.}\ \bibnamefont {Tan}},\ }\href
  {https://doi.org/10.1103/PhysRevLett.123.106402} {\bibfield  {journal}
  {\bibinfo  {journal} {Phys. Rev. Lett.}\ }\textbf {\bibinfo {volume} {123}},\
  \bibinfo {pages} {106402} (\bibinfo {year} {2019})}\BibitemShut {NoStop}%
\bibitem [{\citenamefont {Ji}\ \emph {et~al.}(2022)\citenamefont {Ji},
  \citenamefont {Luo}, \citenamefont {Chen}, \citenamefont {Feng},
  \citenamefont {Hao}, \citenamefont {Liu}, \citenamefont {Zhang},
  \citenamefont {Liu}, \citenamefont {Wang}, \citenamefont {Tan},\ and\
  \citenamefont {Lai}}]{Ji2022PRB}%
  \BibitemOpen
  \bibfield  {author} {\bibinfo {author} {\bibfnamefont {X.}~\bibnamefont
  {Ji}}, \bibinfo {author} {\bibfnamefont {X.}~\bibnamefont {Luo}}, \bibinfo
  {author} {\bibfnamefont {Q.}~\bibnamefont {Chen}}, \bibinfo {author}
  {\bibfnamefont {W.}~\bibnamefont {Feng}}, \bibinfo {author} {\bibfnamefont
  {Q.}~\bibnamefont {Hao}}, \bibinfo {author} {\bibfnamefont {Q.}~\bibnamefont
  {Liu}}, \bibinfo {author} {\bibfnamefont {Y.}~\bibnamefont {Zhang}}, \bibinfo
  {author} {\bibfnamefont {Y.}~\bibnamefont {Liu}}, \bibinfo {author}
  {\bibfnamefont {W.}~\bibnamefont {Wang}}, \bibinfo {author} {\bibfnamefont
  {S.}~\bibnamefont {Tan}},\ and\ \bibinfo {author} {\bibfnamefont
  {X.}~\bibnamefont {Lai}},\ }\href
  {https://doi.org/10.1103/PhysRevB.106.125120} {\bibfield  {journal} {\bibinfo
   {journal} {Phys. Rev. B}\ }\textbf {\bibinfo {volume} {106}},\ \bibinfo
  {pages} {125120} (\bibinfo {year} {2022})}\BibitemShut {NoStop}%
\bibitem [{\citenamefont {Wu}(2017)}]{Broholm2019}%
  \BibitemOpen
  \bibfield  {author} {\bibinfo {author} {\bibfnamefont {S.}~\bibnamefont
  {Wu}},\ }\href@noop {} {\emph {\bibinfo {title} {Magnetism near the Electron
  Localization Transition. Ph. D Thesis}}}\ (\bibinfo  {publisher} {Johns
  Hopkins University},\ \bibinfo {year} {2017})\BibitemShut {NoStop}%
\bibitem [{\citenamefont {Neill}\ \emph {et~al.}(2021)\citenamefont {Neill},
  \citenamefont {Schmehr}, \citenamefont {Keen}, \citenamefont {Cairns},
  \citenamefont {Sokolov}, \citenamefont {Hermann}, \citenamefont {Wermeille},
  \citenamefont {Manuel}, \citenamefont {Kr{\"u}ger},\ and\ \citenamefont
  {Huxley}}]{Neill2021}%
  \BibitemOpen
  \bibfield  {author} {\bibinfo {author} {\bibfnamefont {C.~D.~O.}\
  \bibnamefont {Neill}}, \bibinfo {author} {\bibfnamefont {J.~L.}\ \bibnamefont
  {Schmehr}}, \bibinfo {author} {\bibfnamefont {H.~D.~J.}\ \bibnamefont
  {Keen}}, \bibinfo {author} {\bibfnamefont {L.~P.}\ \bibnamefont {Cairns}},
  \bibinfo {author} {\bibfnamefont {D.~A.}\ \bibnamefont {Sokolov}}, \bibinfo
  {author} {\bibfnamefont {A.}~\bibnamefont {Hermann}}, \bibinfo {author}
  {\bibfnamefont {D.}~\bibnamefont {Wermeille}}, \bibinfo {author}
  {\bibfnamefont {P.}~\bibnamefont {Manuel}}, \bibinfo {author} {\bibfnamefont
  {F.}~\bibnamefont {Kr{\"u}ger}},\ and\ \bibinfo {author} {\bibfnamefont
  {A.~D.}\ \bibnamefont {Huxley}},\ }\href
  {https://doi.org/10.1073/pnas.2102687118} {\bibfield  {journal} {\bibinfo
  {journal} {Proceedings of the National Academy of Sciences}\ }\textbf
  {\bibinfo {volume} {118}},\ \bibinfo {pages} {e2102687118} (\bibinfo {year}
  {2021})}\BibitemShut {NoStop}%
\bibitem [{\citenamefont {Kroha}(2017)}]{Kroha2017Interplay}%
  \BibitemOpen
  \bibfield  {author} {\bibinfo {author} {\bibfnamefont {J.}~\bibnamefont
  {Kroha}},\ }\href {https://arxiv.org/abs/1710.00192} {\bibfield  {journal}
  {\bibinfo  {journal} {arXiv preprint arXiv:1710.00192}\ } (\bibinfo {year}
  {2017})}\BibitemShut {NoStop}%
\bibitem [{\citenamefont {Shishido}\ \emph {et~al.}(2002)\citenamefont
  {Shishido}, \citenamefont {Settai}, \citenamefont {Aoki}, \citenamefont
  {Ikeda}, \citenamefont {Nakawaki}, \citenamefont {Nakamura}, \citenamefont
  {Iizuka}, \citenamefont {Inada}, \citenamefont {Sugiyama}, \citenamefont
  {Takeuchi}, \citenamefont {Kindo}, \citenamefont {Kobayashi}, \citenamefont
  {Haga}, \citenamefont {Harima}, \citenamefont {Aoki}, \citenamefont {Namiki},
  \citenamefont {Sato},\ and\ \citenamefont {Ōnuki}}]{shishido2002JPSJ}%
  \BibitemOpen
  \bibfield  {author} {\bibinfo {author} {\bibfnamefont {H.}~\bibnamefont
  {Shishido}}, \bibinfo {author} {\bibfnamefont {R.}~\bibnamefont {Settai}},
  \bibinfo {author} {\bibfnamefont {D.}~\bibnamefont {Aoki}}, \bibinfo {author}
  {\bibfnamefont {S.}~\bibnamefont {Ikeda}}, \bibinfo {author} {\bibfnamefont
  {H.}~\bibnamefont {Nakawaki}}, \bibinfo {author} {\bibfnamefont
  {N.}~\bibnamefont {Nakamura}}, \bibinfo {author} {\bibfnamefont
  {T.}~\bibnamefont {Iizuka}}, \bibinfo {author} {\bibfnamefont
  {Y.}~\bibnamefont {Inada}}, \bibinfo {author} {\bibfnamefont
  {K.}~\bibnamefont {Sugiyama}}, \bibinfo {author} {\bibfnamefont
  {T.}~\bibnamefont {Takeuchi}}, \bibinfo {author} {\bibfnamefont
  {K.}~\bibnamefont {Kindo}}, \bibinfo {author} {\bibfnamefont {T.~C.}\
  \bibnamefont {Kobayashi}}, \bibinfo {author} {\bibfnamefont {Y.}~\bibnamefont
  {Haga}}, \bibinfo {author} {\bibfnamefont {H.}~\bibnamefont {Harima}},
  \bibinfo {author} {\bibfnamefont {Y.}~\bibnamefont {Aoki}}, \bibinfo {author}
  {\bibfnamefont {T.}~\bibnamefont {Namiki}}, \bibinfo {author} {\bibfnamefont
  {H.}~\bibnamefont {Sato}},\ and\ \bibinfo {author} {\bibfnamefont
  {Y.}~\bibnamefont {Ōnuki}},\ }\href {https://doi.org/10.1143/JPSJ.71.162}
  {\bibfield  {journal} {\bibinfo  {journal} {Journal of the Physical Society
  of Japan}\ }\textbf {\bibinfo {volume} {71}},\ \bibinfo {pages} {162}
  (\bibinfo {year} {2002})}\BibitemShut {NoStop}%
\bibitem [{\citenamefont {Harrison}\ \emph {et~al.}(2004)\citenamefont
  {Harrison}, \citenamefont {Alver}, \citenamefont {Goodrich}, \citenamefont
  {Vekhter}, \citenamefont {Sarrao}, \citenamefont {Pagliuso}, \citenamefont
  {Moreno}, \citenamefont {Balicas}, \citenamefont {Fisk}, \citenamefont
  {Hall}, \citenamefont {Macaluso},\ and\ \citenamefont
  {Chan}}]{Harrison2004PRL}%
  \BibitemOpen
  \bibfield  {author} {\bibinfo {author} {\bibfnamefont {N.}~\bibnamefont
  {Harrison}}, \bibinfo {author} {\bibfnamefont {U.}~\bibnamefont {Alver}},
  \bibinfo {author} {\bibfnamefont {R.~G.}\ \bibnamefont {Goodrich}}, \bibinfo
  {author} {\bibfnamefont {I.}~\bibnamefont {Vekhter}}, \bibinfo {author}
  {\bibfnamefont {J.~L.}\ \bibnamefont {Sarrao}}, \bibinfo {author}
  {\bibfnamefont {P.~G.}\ \bibnamefont {Pagliuso}}, \bibinfo {author}
  {\bibfnamefont {N.~O.}\ \bibnamefont {Moreno}}, \bibinfo {author}
  {\bibfnamefont {L.}~\bibnamefont {Balicas}}, \bibinfo {author} {\bibfnamefont
  {Z.}~\bibnamefont {Fisk}}, \bibinfo {author} {\bibfnamefont {D.}~\bibnamefont
  {Hall}}, \bibinfo {author} {\bibfnamefont {R.~T.}\ \bibnamefont {Macaluso}},\
  and\ \bibinfo {author} {\bibfnamefont {J.~Y.}\ \bibnamefont {Chan}},\ }\href
  {https://doi.org/10.1103/PhysRevLett.93.186405} {\bibfield  {journal}
  {\bibinfo  {journal} {Phys. Rev. Lett.}\ }\textbf {\bibinfo {volume} {93}},\
  \bibinfo {pages} {186405} (\bibinfo {year} {2004})}\BibitemShut {NoStop}%
\bibitem [{\citenamefont {Chen}\ \emph {et~al.}(2018)\citenamefont {Chen},
  \citenamefont {Xu}, \citenamefont {Niu}, \citenamefont {Peng}, \citenamefont
  {Xu}, \citenamefont {Wen}, \citenamefont {Liu}, \citenamefont {Shu},
  \citenamefont {Tan}, \citenamefont {Lai}, \citenamefont {Zhang},
  \citenamefont {Lee}, \citenamefont {Strocov}, \citenamefont {Bisti},
  \citenamefont {Dudin}, \citenamefont {Zhu}, \citenamefont {Yuan},
  \citenamefont {Kirchner},\ and\ \citenamefont {Feng}}]{Chen2018}%
  \BibitemOpen
  \bibfield  {author} {\bibinfo {author} {\bibfnamefont {Q.~Y.}\ \bibnamefont
  {Chen}}, \bibinfo {author} {\bibfnamefont {D.~F.}\ \bibnamefont {Xu}},
  \bibinfo {author} {\bibfnamefont {X.~H.}\ \bibnamefont {Niu}}, \bibinfo
  {author} {\bibfnamefont {R.}~\bibnamefont {Peng}}, \bibinfo {author}
  {\bibfnamefont {H.~C.}\ \bibnamefont {Xu}}, \bibinfo {author} {\bibfnamefont
  {C.~H.~P.}\ \bibnamefont {Wen}}, \bibinfo {author} {\bibfnamefont
  {X.}~\bibnamefont {Liu}}, \bibinfo {author} {\bibfnamefont {L.}~\bibnamefont
  {Shu}}, \bibinfo {author} {\bibfnamefont {S.~Y.}\ \bibnamefont {Tan}},
  \bibinfo {author} {\bibfnamefont {X.~C.}\ \bibnamefont {Lai}}, \bibinfo
  {author} {\bibfnamefont {Y.~J.}\ \bibnamefont {Zhang}}, \bibinfo {author}
  {\bibfnamefont {H.}~\bibnamefont {Lee}}, \bibinfo {author} {\bibfnamefont
  {V.~N.}\ \bibnamefont {Strocov}}, \bibinfo {author} {\bibfnamefont
  {F.}~\bibnamefont {Bisti}}, \bibinfo {author} {\bibfnamefont
  {P.}~\bibnamefont {Dudin}}, \bibinfo {author} {\bibfnamefont {J.-X.}\
  \bibnamefont {Zhu}}, \bibinfo {author} {\bibfnamefont {H.~Q.}\ \bibnamefont
  {Yuan}}, \bibinfo {author} {\bibfnamefont {S.}~\bibnamefont {Kirchner}},\
  and\ \bibinfo {author} {\bibfnamefont {D.~L.}\ \bibnamefont {Feng}},\ }\href
  {https://doi.org/10.1103/PhysRevLett.120.066403} {\bibfield  {journal}
  {\bibinfo  {journal} {Phys. Rev. Lett.}\ }\textbf {\bibinfo {volume} {120}},\
  \bibinfo {pages} {066403} (\bibinfo {year} {2018})}\BibitemShut {NoStop}%
\bibitem [{\citenamefont {Okuda}\ \emph {et~al.}(2007)\citenamefont {Okuda},
  \citenamefont {Miyauchi}, \citenamefont {Ida}, \citenamefont {Takeda},
  \citenamefont {Tonohiro}, \citenamefont {Oduchi}, \citenamefont {Yamada},
  \citenamefont {Duc~Dung}, \citenamefont {D.~Matsuda}, \citenamefont {Haga},
  \citenamefont {Takeuchi}, \citenamefont {Hagiwara}, \citenamefont {Kindo},
  \citenamefont {Harima}, \citenamefont {Sugiyama}, \citenamefont {Settai},\
  and\ \citenamefont {\={O}nuki}}]{Okuda2007}%
  \BibitemOpen
  \bibfield  {author} {\bibinfo {author} {\bibfnamefont {Y.}~\bibnamefont
  {Okuda}}, \bibinfo {author} {\bibfnamefont {Y.}~\bibnamefont {Miyauchi}},
  \bibinfo {author} {\bibfnamefont {Y.}~\bibnamefont {Ida}}, \bibinfo {author}
  {\bibfnamefont {Y.}~\bibnamefont {Takeda}}, \bibinfo {author} {\bibfnamefont
  {C.}~\bibnamefont {Tonohiro}}, \bibinfo {author} {\bibfnamefont
  {Y.}~\bibnamefont {Oduchi}}, \bibinfo {author} {\bibfnamefont
  {T.}~\bibnamefont {Yamada}}, \bibinfo {author} {\bibfnamefont
  {N.}~\bibnamefont {Duc~Dung}}, \bibinfo {author} {\bibfnamefont
  {T.}~\bibnamefont {D.~Matsuda}}, \bibinfo {author} {\bibfnamefont
  {Y.}~\bibnamefont {Haga}}, \bibinfo {author} {\bibfnamefont {T.}~\bibnamefont
  {Takeuchi}}, \bibinfo {author} {\bibfnamefont {M.}~\bibnamefont {Hagiwara}},
  \bibinfo {author} {\bibfnamefont {K.}~\bibnamefont {Kindo}}, \bibinfo
  {author} {\bibfnamefont {H.}~\bibnamefont {Harima}}, \bibinfo {author}
  {\bibfnamefont {K.}~\bibnamefont {Sugiyama}}, \bibinfo {author}
  {\bibfnamefont {R.}~\bibnamefont {Settai}},\ and\ \bibinfo {author}
  {\bibfnamefont {Y.}~\bibnamefont {\={O}nuki}},\ }\href
  {https://doi.org/10.1143/JPSJ.76.044708} {\bibfield  {journal} {\bibinfo
  {journal} {J. Phys. Soc. Jpn.}\ }\textbf {\bibinfo {volume} {76}},\ \bibinfo
  {pages} {044708} (\bibinfo {year} {2007})}\BibitemShut {NoStop}%
\bibitem [{\citenamefont {Terashima}\ \emph {et~al.}(2008)\citenamefont
  {Terashima}, \citenamefont {Kimata}, \citenamefont {Uji}, \citenamefont
  {Sugawara}, \citenamefont {Kimura}, \citenamefont {Aoki},\ and\ \citenamefont
  {Harima}}]{Terashima2008}%
  \BibitemOpen
  \bibfield  {author} {\bibinfo {author} {\bibfnamefont {T.}~\bibnamefont
  {Terashima}}, \bibinfo {author} {\bibfnamefont {M.}~\bibnamefont {Kimata}},
  \bibinfo {author} {\bibfnamefont {S.}~\bibnamefont {Uji}}, \bibinfo {author}
  {\bibfnamefont {T.}~\bibnamefont {Sugawara}}, \bibinfo {author}
  {\bibfnamefont {N.}~\bibnamefont {Kimura}}, \bibinfo {author} {\bibfnamefont
  {H.}~\bibnamefont {Aoki}},\ and\ \bibinfo {author} {\bibfnamefont
  {H.}~\bibnamefont {Harima}},\ }\href
  {https://doi.org/10.1103/PhysRevB.78.205107} {\bibfield  {journal} {\bibinfo
  {journal} {Phys. Rev. B}\ }\textbf {\bibinfo {volume} {78}},\ \bibinfo
  {pages} {205107} (\bibinfo {year} {2008})}\BibitemShut {NoStop}%
\bibitem [{\citenamefont {Ohkochi}\ \emph {et~al.}(2009)\citenamefont
  {Ohkochi}, \citenamefont {Toshimitsu}, \citenamefont {Yamagami},
  \citenamefont {Fujimori}, \citenamefont {Yasui}, \citenamefont {Takeda},
  \citenamefont {Okane}, \citenamefont {Saitoh}, \citenamefont {Fujimori},
  \citenamefont {Miyauchi}, \citenamefont {Okuda}, \citenamefont {Settai},\
  and\ \citenamefont {\={O}nuki}}]{Ohkochi2009JPSJ}%
  \BibitemOpen
  \bibfield  {author} {\bibinfo {author} {\bibfnamefont {T.}~\bibnamefont
  {Ohkochi}}, \bibinfo {author} {\bibfnamefont {T.}~\bibnamefont {Toshimitsu}},
  \bibinfo {author} {\bibfnamefont {H.}~\bibnamefont {Yamagami}}, \bibinfo
  {author} {\bibfnamefont {S.}~\bibnamefont {Fujimori}}, \bibinfo {author}
  {\bibfnamefont {A.}~\bibnamefont {Yasui}}, \bibinfo {author} {\bibfnamefont
  {Y.}~\bibnamefont {Takeda}}, \bibinfo {author} {\bibfnamefont
  {T.}~\bibnamefont {Okane}}, \bibinfo {author} {\bibfnamefont
  {Y.}~\bibnamefont {Saitoh}}, \bibinfo {author} {\bibfnamefont
  {A.}~\bibnamefont {Fujimori}}, \bibinfo {author} {\bibfnamefont
  {Y.}~\bibnamefont {Miyauchi}}, \bibinfo {author} {\bibfnamefont
  {Y.}~\bibnamefont {Okuda}}, \bibinfo {author} {\bibfnamefont
  {R.}~\bibnamefont {Settai}},\ and\ \bibinfo {author} {\bibfnamefont
  {Y.}~\bibnamefont {\={O}nuki}},\ }\href
  {https://doi.org/10.1143/JPSJ.78.084802} {\bibfield  {journal} {\bibinfo
  {journal} {J. Phys. Soc. Jpn.}\ }\textbf {\bibinfo {volume} {78}},\ \bibinfo
  {pages} {084802} (\bibinfo {year} {2009})}\BibitemShut {NoStop}%
\bibitem [{\citenamefont {Smidman}\ \emph {et~al.}(2015)\citenamefont
  {Smidman}, \citenamefont {Adroja}, \citenamefont {Goremychkin}, \citenamefont
  {Lees}, \citenamefont {Paul},\ and\ \citenamefont
  {Balakrishnan}}]{Smidman2015PRBCeRuSi3}%
  \BibitemOpen
  \bibfield  {author} {\bibinfo {author} {\bibfnamefont {M.}~\bibnamefont
  {Smidman}}, \bibinfo {author} {\bibfnamefont {D.~T.}\ \bibnamefont {Adroja}},
  \bibinfo {author} {\bibfnamefont {E.~A.}\ \bibnamefont {Goremychkin}},
  \bibinfo {author} {\bibfnamefont {M.~R.}\ \bibnamefont {Lees}}, \bibinfo
  {author} {\bibfnamefont {D.~M.}\ \bibnamefont {Paul}},\ and\ \bibinfo
  {author} {\bibfnamefont {G.}~\bibnamefont {Balakrishnan}},\ }\href
  {https://doi.org/10.1103/PhysRevB.91.064419} {\bibfield  {journal} {\bibinfo
  {journal} {Phys. Rev. B}\ }\textbf {\bibinfo {volume} {91}},\ \bibinfo
  {pages} {064419} (\bibinfo {year} {2015})}\BibitemShut {NoStop}%
\bibitem [{\citenamefont {P{\'a}sztorov{\'a}}\ \emph
  {et~al.}(2019)\citenamefont {P{\'a}sztorov{\'a}}, \citenamefont {Howell},
  \citenamefont {Songvilay}, \citenamefont {Sarte}, \citenamefont
  {Rodriguez-Rivera}, \citenamefont {Ar{\'e}valo-L{\'o}pez}, \citenamefont
  {Schmalzl}, \citenamefont {Scheneidewind}, \citenamefont {Dunsiger},
  \citenamefont {Singh}, \citenamefont {Petrovic}, \citenamefont {Hu},\ and\
  \citenamefont {Stock}}]{pasztorova2015}%
  \BibitemOpen
  \bibfield  {author} {\bibinfo {author} {\bibfnamefont {J.}~\bibnamefont
  {P{\'a}sztorov{\'a}}}, \bibinfo {author} {\bibfnamefont {A.}~\bibnamefont
  {Howell}}, \bibinfo {author} {\bibfnamefont {M.}~\bibnamefont {Songvilay}},
  \bibinfo {author} {\bibfnamefont {P.}~\bibnamefont {Sarte}}, \bibinfo
  {author} {\bibfnamefont {J.}~\bibnamefont {Rodriguez-Rivera}}, \bibinfo
  {author} {\bibfnamefont {A.}~\bibnamefont {Ar{\'e}valo-L{\'o}pez}}, \bibinfo
  {author} {\bibfnamefont {K.}~\bibnamefont {Schmalzl}}, \bibinfo {author}
  {\bibfnamefont {A.}~\bibnamefont {Scheneidewind}}, \bibinfo {author}
  {\bibfnamefont {S.}~\bibnamefont {Dunsiger}}, \bibinfo {author}
  {\bibfnamefont {D.}~\bibnamefont {Singh}}, \bibinfo {author} {\bibfnamefont
  {C.}~\bibnamefont {Petrovic}}, \bibinfo {author} {\bibfnamefont
  {R.}~\bibnamefont {Hu}},\ and\ \bibinfo {author} {\bibfnamefont
  {C.}~\bibnamefont {Stock}},\ }\href@noop {} {\bibfield  {journal} {\bibinfo
  {journal} {Physical Review B}\ }\textbf {\bibinfo {volume} {99}},\ \bibinfo
  {pages} {125144} (\bibinfo {year} {2019})}\BibitemShut {NoStop}%
\end{thebibliography}%

\end{document}